\documentclass[fleqn,usenatbib]{mnras}
\usepackage{newtxtext,newtxmath}
\usepackage[T1]{fontenc}
\usepackage{ae,aecompl}

\usepackage{graphicx}	
\usepackage{amsmath}
\usepackage{url}
\usepackage{color}

\usepackage[normalem]{ulem}

\renewcommand{\d}{{\rm d}}
\renewcommand{\bar}[1]{\overline{#1}}

\newcommand{\new}[1]{\textcolor{black}{#1}}
\newcommand{\neww}[1]{\textcolor{black}{#1}}

\title[A geometrical model for flares in gamma-ray burst afterglows]{Flares in gamma-ray burst X-ray afterglows as prompt emission from slightly misaligned structured jets}

\author[Duque et al.]{
Rapha\"{e}l Duque$^{1, 2}$\thanks{E-mail: duque@physik.uni-frankfurt.de},
Paz Beniamini$^{3, 4, 5}$,
Fr\'{e}d\'{e}ric Daigne$^{1}$
and Robert Mochkovitch$^{1}$
\\
$^{1}$Sorbonne Universit\'{e}, CNRS, UMR 7095, Institut d'Astrophysique de Paris, 98 bis boulevard Arago, 75014 Paris, France\\
$^{2}$Institut für Theoretische Physik, Goethe Universität Frankfurt am Main, D-60323 Frankfurt am Main, Germany\\
$^{3}$Department of Natural Sciences, Open University of Israel, 1 University Road, 43107 Ra'anana, Israel\\
$^{4}$Astrophysics Research Center of the Open university (ARCO), The Open University of Israel, P.O Box 808, Ra’anana 43537, Israel\\
$^{5}$Division of Physics, Mathematics and Astronomy, California Institute of Technology, Pasadena, CA 91125, USA}

\date{Accepted XXX. Received YYY; in original form ZZZ}

\pubyear{2021}

\begin{document}
\label{firstpage}
\pagerange{\pageref{firstpage}--\pageref{lastpage}}
\maketitle

\begin{abstract}
We develop a model to explain the flaring activity in gamma-ray burst X-ray afterglows within the framework of slightly misaligned observers to structured jets. We suggest that flares could be the manifestation of prompt dissipation within the core of the jet, appearing to a misaligned observer in the X-ray band because of less favorable Doppler boosting. These flares appear during the afterglow phase because of core--observer light travel delays. In this picture, the prompt emission recorded by this observer comes from material along their line of sight, in the lateral structure of the jet, outside the jet's core. We start by laying down the basic analytical framework to determine the flares characteristics as a function of those of the gamma-ray pulse an aligned observer would see. We show that there is viable parameter space to explain flares with typical observing times and luminosities. We then analytically explore this model, showing that it naturally produces flares with small aspect ratios, as observed. We perform fits of our model to two \textit{Swift}/XRT flares representing two different types of morphology, to show that our model can capture both. The ejection time of the core jet material responsible of the flare is a critical parameter. While it always remains small compared to the observed time of the flare, confirming that our model does not require very late central engine activity, late ejection times are strongly favored, sometimes larger than the observed duration of the parent gamma-ray burst's prompt emission as measured by $T_{90}$.
\end{abstract}




\section{Introduction}
\label{sec:intro}

Flares are sudden rebrightenings in the afterglow phases of gamma-ray bursts (GRBs), observed primarily in the X-ray band \citep{2005Sci...309.1833B,NKGPG+2006}, though also in the optical \citep{2012ApJ...758...27L,2013ApJ...774....2S,2017ApJ...844...79Y} and radio bands \citep[e.g.][]{2007A&A...473...77M}. Flares occur in around one third of observed GRB afterglows, and most observed flares occur less than 1000~s after the prompt trigger \citep[e.g.,][]{2010MNRAS.406.2113C,2016ApJS..224...20Y}. From the first catalogs of flares in \textit{Swift}/XRT light curves, flares showed the salient feature of having small and tightly distributed \textit{aspect ratios} (i.e., the ratio of their width $\Delta t$ to their arrival time $t$, \citealt{2007ApJ...671.1903C}). It was also noted that flare morphologies were quite diverse---with both fast and slow rising and decay phases---and that they mimicked GRB prompt pulses, with a very similar distribution of rise-to-decay-time ratios \citep{2010MNRAS.406.2113C}. Flares exhibit other features remarkably analogous to prompt pulses such as spectral lags, lag--luminosity correlations and width--energy band relations \citep{2010MNRAS.406.2149M}. \new{Pulse temporal profile and variability timescale analyses of prompt emission and X-ray flares also find common traits for both phenomena \citep{2013ApJ...767L..28S,2015ApJ...801...57G}.}

These temporal and spectral similarities with prompt pulses suggest a common origin for flares and prompt emission. For example, the central engine of GRBs could have a second episode of activity, explaining the delay between prompt emission and X-ray flares \citep[e.g.,][]{2005Sci...309.1833B,FanWei2005,2006ApJ...646..351L,LazzatiPerna2007}. Causes for the engine's delayed restarting can, for example, be fragmentation and accretion of a collapsing star \citep{2005ApJ...630L.113K}, instability-induced variability in accretion around the central object \citep{2006ApJ...636L..29P} or magnetic activity of the young pulsar produced by the merger in non-collapsar GRBs \citep{2006Sci...311.1127D}. However, in these late-engine-activity models, the emission in the X-ray rather than the gamma-ray bands requires an explanation. Furthermore, producing small aspect ratios often requires to tune the second activity's duration to the time of quiescence between the two episodes in an unnatural way. Finally, in some cases, the energy requirements for powering the X-ray flares are often incompatible with the energy available {\it at that time} \citep{BK2016}, from, e.g., a rapidly spinning magnetar---a widely invoked model for GRB central engines. \new{Alternatively, the central engine can have a single episode of ejection, in which case the dissipation of energy in the jet can be delayed, as in delayed magnetic dissipation or internal shocks \citep{2006A&A...455L...5G,2009ApJ...692..133Y,2015ApJ...803...10T,2018A&A...615A..80P}}. Other models for flares suggest an origin distinct from that of prompt dissipation, such as emission from the reverse shock propagating in a stratified ejecta \citep{2017MNRAS.472L..94H,2018MNRAS.474.2813L,2020MNRAS.495.2979A}, Compton up-scattering of photons from the reverse shock when crossing the forward shock \citep{2007ApJ...655..391K} or from the forward shock on a preceding relativistic outflow \citep{2008MNRAS.383.1143P}, or photospheric emission from material moving with modest Lorentz factors compared to that producing the prompt, but ejected at roughly the same time \citep{BK2016}.

\begin{figure}
\includegraphics[width=0.5\textwidth]{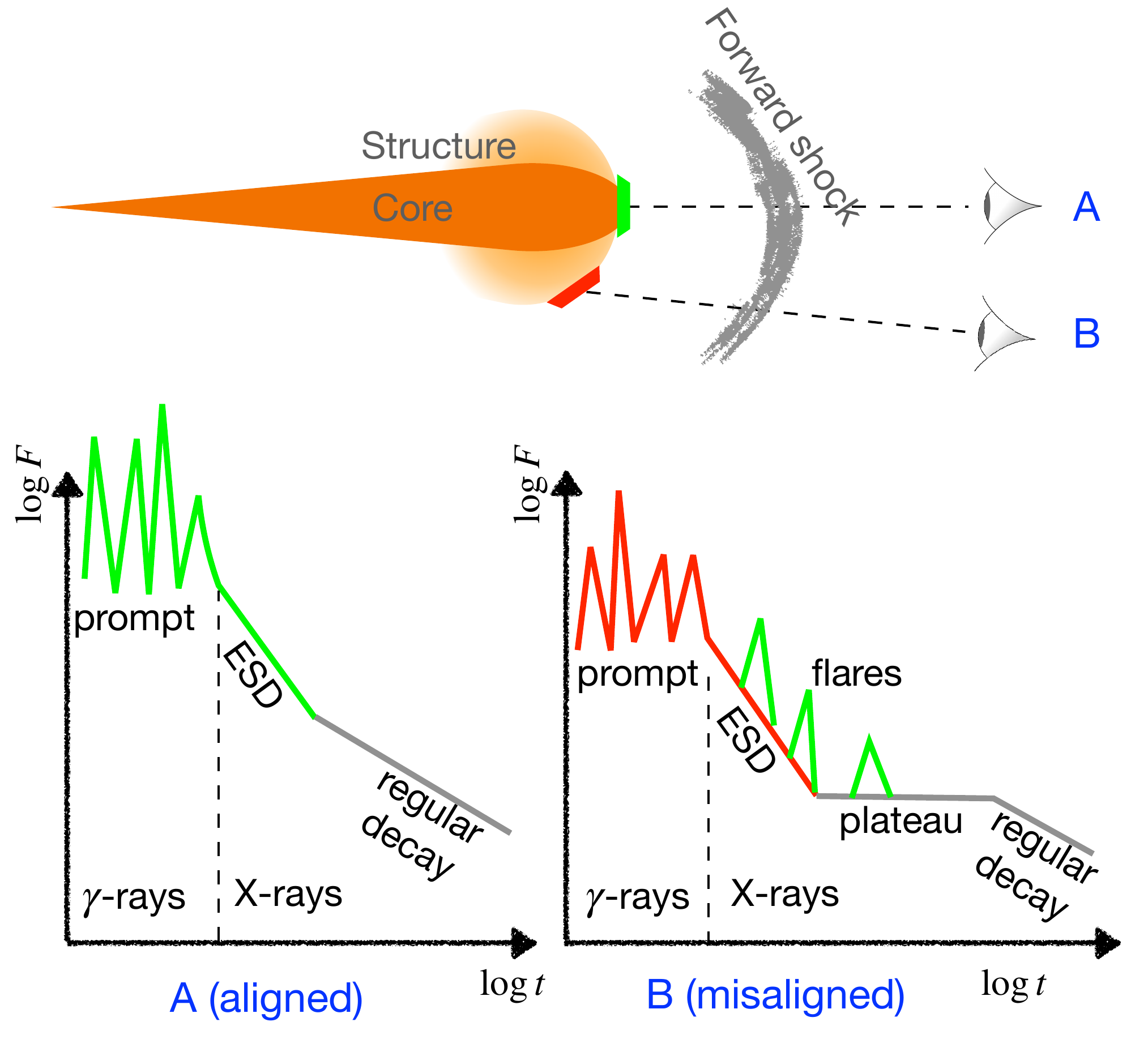}
\caption{Schematic description of our unified picture for flares and plateaus. The colors of the elements of the light curves correspond with the colors of the emitting regions in the jet: core (green) or lateral structure (red). Each component of the prompt and afterglow phases of aligned (A) and misaligned (B) viewers comes from a different region, according to our picture. For the misaligned observer, the emission from the core matter appears as flares in the X-ray band, atop the ESD and the plateau phase.}
\label{fig:schematic}
\end{figure}

In a recent publication \citep{BDDM2020}, we developed a model to explain the emergence of plateaus in some GRB afterglows, another remarkable feature of GRB afterglows consisting of extended phases of near-flat flux evolution occurring before the regular decay of afterglows \citep{NKGPG+2006}. This model, initially suggested by \citet{EG2006}, is based on a setup where the GRB jet possesses angular structure and the observer is slightly misaligned with the GRB jet, to within a few times the jet's core half-opening angle, i.e., $\theta_v \lesssim 2 \theta_j$. It is indeed likely that this regime of viewing angles is that assumed when detecting GRBs observed at cosmological distances \citep{BN2019}. In this model, the shallow flux evolution of the plateau is produced by the fact that any region in the decelerating jet is only revealed to the observer when this region has slowed down enough for the line of sight to be included in the material's beaming cone. Thus, as the jet decelerates, the observer progressively discovers material closer to the jet's core. Because of the jet structure, this material is intrinsically brighter but less boosted because it is pointed further away from the observer. This can lead to a shallow evolution of the total afterglow flux. In \citet{BDDM2020}, we derived analytically the duration and flux of plateaus expected in this picture as a function of the jet structure and the observer's viewing angle. We showed how established correlations between plateau duration and flux level and the parent GRB's prompt properties naturally arise in this model.

\new{Recently, GRB jet structure has become an increasingly important ingredient in GRB afterglow modeling \citep[e.g.,][]{2020ApJ...893...88O,2020A&A...641A..61A,BGG2020,2021MNRAS.506.4163L,2021MNRAS.501.5746T}, especially since the historical insight provided by the multi-messenger analysis of the outflow from GW170817 \citep{MDGNH+2018,GSPGY+2019}}. In the present paper, we set out to interpret flares in GRB afterglows within the same physical setup as our above-mentioned plateau model: slightly misaligned lines of sight to a structured jet. Acknowledging the aforementioned similarities between X-ray flares and GRB prompt pulses, we also posit a common origin for the two. However, we will explain the delayed occurrence of the flares not by their delayed emission, but rather by the light travel time between the flare production site within the core and the misaligned observer: we suggest that flares in GRB X-ray afterglows are the manifestation of prompt dissipation in the core of the jet, as seen from slightly off-axis lines of sight. Because of relativistic effects, this radiation appears delayed, dimmer and downshifted in energy. In other words, X-ray flares are \textit{deboosted} versions of gamma-ray pulses from prompt energy dissipation in the core.

We present our unified picture for plateaus and flares in Fig.~\ref{fig:schematic}: For an aligned viewer (A), the prompt emission comes from the core jet shining in gamma-rays (green), and the afterglow phase contains the early steep decay and radiation from the decelerating forward shock (grey); all other jet regions are too weak and not boosted enough to contribute to the aligned observer's signal. For a misaligned observer (B), the prompt emission and early steep decay come from the material down their line of sight (red). Progressively the structured jet decelerates, giving rise to the plateau phase (grey). In the mean time, prompt photons from the core (green) travel to the observer, and reach them as X-ray flares, i.e., deboosted and dimmer than photons that would have reached the aligned observer.

Note that we consider a single central engine activity episode, and the delay in flare occurrence is a geometrical effect. We anticipate Sec.~\ref{sec:outline} in mentioning that the arrival time for flares in such a picture is bounded to $t_{\rm flares} \lesssim 1000$~s after the prompt trigger. We are therefore dealing with \textit{early flares}. Remarkably, these early flares seem to constitute a distinct (and largely statistically dominant, \citealt{2014ApJ...788...30S}) class of flares, as shown by their distinctive post-peak decay slopes and distribution of flare-to-continuum contrasts ($\Delta F / F$) with respect to late flares ($t_{\rm flare} \gtrsim 1000$~s, \citealt{2011MNRAS.410.1064M,2011A&A...526A..27B}). \new{While the dichotomy in $\Delta F/F$ could result from the late flares most often occurring during the plateau phase of the afterglow and the early flares during the early steep decay, the dichotomy in decay slopes} suggests a different origin for these two classes, and our model's natural restriction to the early class is a further motivation to explore this picture.

\new{We outline this new framework for flares and make a first exploration of parameter space in Sec.~\ref{sec:outline}. We study the conditions for flares to appear above the underlying afterglow continuum in Sec.~\ref{sec:visi}, where we also expose a feature of our model by which flares that rise above the afterglow continuum tend be thinner. In Sec.~\ref{sec:morpho}, we confront our model to \textit{Swift} X-ray data by making fits to light curves of actual GRB afterglows with flares. We summarize our results and discuss the underlying prompt dissipation mechanism and consequences of our model in Sec.~\ref{sec:discussion}.}

\section{Model outline and basic properties}
\label{sec:outline}

\begin{figure}
\includegraphics[width=\linewidth]{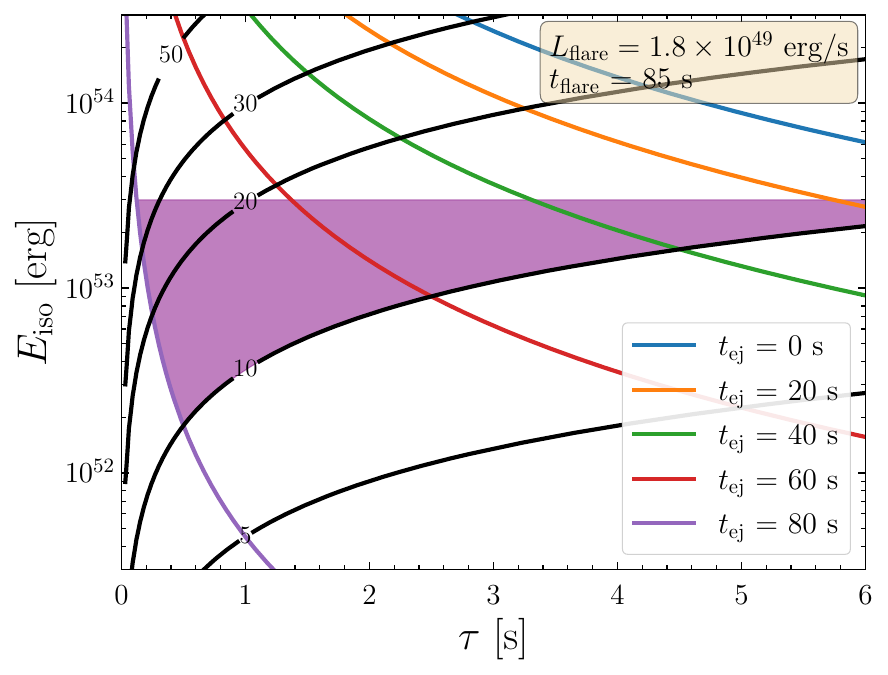}
\caption{Misaligned structured-jet flare model parameter space exploration to reproduce the median XRT flare. The colored lines represent the $E_{\rm iso}$ energy required to produce a typical flare peaking at $t_{\rm flare} = 85~{\rm s}$ and with peak luminosity $L_{\rm flare} = 1.8\times 10^{49}~{\rm erg/s}$ as a function of the shell's geometrical timescale $\tau$ and for different ejections times from 0 to 80 s, an estimate of GRB central engine activity duration. The black lines are contours of the corresponding required $\mathcal{S}$-factor, values are indicated on the lines. The region in this parameter space consistent with the off-axis geometrical setup of our model and with expected ranges for shell dissipated energies and GRB pulse durations is marked in purple. We find that there is available parameter space for our model to explain the typical flare.}
\label{fig:2dfit}
\end{figure}

We consider a shell of ultra-relativistic matter with Lorentz factor $\Gamma$ ejected at a time $t_{\rm ej}$ from the central engine, within the core of the jet. At an emission time $t_e$, this shell reaches a dissipation radius $R_e = \beta \left(t_e-t_{\rm ej}\right)$ and radiates energy which, for a core-aligned observer appears as gamma rays. For this aligned observer, this radiation is observed at time:
\begin{equation}
t_{\rm on} = t_{\rm ej} +\frac{1-\beta}{\beta}\frac{R_e}{c}
\label{eq:1}
\end{equation}
where $\beta \sim 1$ is the shell's velocity.

We now consider a misaligned observer, lying at a viewing angle $\theta_v$ from the jet's axis, with $\theta_v > \theta_j$, where $\theta_j$ is the core's half-opening angle. For this observer, the first photons from this shell's radiation arrive at a time $t_{\rm off} = t_{\rm ej} + \frac{1 - \beta \cos (\theta_v - \theta_j)}{\beta}\frac{R_e}{c}$.
Stated differently:
\begin{equation}
t_{\rm off} = \mathcal{S} t_{\rm on} - (\mathcal{S} - 1) t_{\rm ej}
\label{eq:2}
\end{equation}
where we have denoted:
\begin{equation}
\mathcal{S} \equiv \frac{{1-\beta\,{\rm cos}\,(\theta_{v}-\theta_{ j})}}{1-\beta},
\end{equation}
which we will refer to as the \textit{stretch factor}. This factor is the ratio of the Doppler boosts between the aligned and misaligned observers. \new{In the ultra-relativistic limit ($\Gamma\gg 1$) and for slightly off-axis observers ($\theta_v-\theta_j \ll 1$), the stretch factor reduces to $\mathcal{S}\simeq 1+\Gamma^2\left(\theta_v-\theta_j\right)^2$.}

Furthermore, the peak energy of the detected spectrum transforms as $\mathcal{S}^{-1}$ between the two observers' rest frames:
\begin{equation}
e_{\rm off} = \mathcal{S}^{-1} e_{\rm on}
\end{equation}

Finally, considering a slightly misaligned line of sight ($\theta_j < \theta_v \lesssim  2 \theta_j$)
we show in Appendix~\ref{sec:B} that the peak bolometric luminosity of the detected radiation transforms according to (Eq.~\ref{eq:B:lum}):
\begin{equation}
L_{\rm off} = f_{\rm geo} \mathcal{S}^{-3} L_{\rm on}
\label{eq:l}
\end{equation}
where $f_{\rm geo} \le 1/2$ is a numerical factor accounting for the transverse angular size of the core jet as seen from the slightly off-axis line of sight. For an observer on the edge of the core, $f_{\rm geo} = 1/2$, and $f_{\rm geo}$ decreases with larger viewing angle. \new{In \citet{BDDM2020}, we considered the observer to be placed at $\theta_v/\theta_j \sim 1.3$ 
\new{to illustrate the results of our} plateau model. In our numerical exploration of the present model (Sec.~\ref{sec:hle}), we shall adopt this same geometric configuration. In this case, we find that Eq.~\ref{eq:l} is best reproduced with $f_{\rm geo} \sim 0.14$.}

These relations show that what is seen as a prompt pulse by an observer aligned with the jet core transforms, for a slightly misaligned observer, into a signal that is delayed, softer and less luminous. In this picture, we will use the term \textit{pulse} for the signal detected by a core-aligned observer, and \textit{flare} for the deboosted emission a misaligned observer detects. A gamma-ray pulse turns into an X-ray flare when seen by an off-axis observer.

For example, let us consider $\Gamma=150$ and $\theta_{v}-\theta_{j}=0.03\,{\rm rad}\sim 5/\Gamma$, a viewing angle that could typically lead to plateau behavior in the afterglow once the structure decelerates, according to the misaligned-observer interpretation developed in \citet[][see their Fig.~1]{BDDM2020}. This configuration leads to $\mathcal{S} = 21$. We take the following typical GRB pulse characteristics: a pulse observed at $t_{\rm on}=8\,{\rm s}$, coming from a shell ejected at $t_{\rm ej}=3\,{\rm s}$ with peak bolometric luminosity $L_{\rm on}=5\times10^{52}\,{\rm erg/s}$ and peak energy $e_{\rm on}=300\,{\rm keV}$. On the slightly misaligned line of sight, such a pulse would appear as an afterglow rebrightening occurring at $t=110\,{\rm s}$ with a peak energy of $e_{\rm off}=14\,{\rm keV}$, in the X-ray band. The X-ray luminosity---or bolometric, as the signal peaks in the X-rays---of this flare is $L_{\rm off}=3\times 10^{48}\,{\rm erg/s}$. It seems therefore that this mechanism can explain the flaring activity in GRB afterglows.

Let us generalize this order-of-magnitude estimate. We denote by $L_{\rm flare}$ the peak bolometric luminosity of an observed flare and $t_{\rm flare}$ its peak time. We seek conditions under which this flare can be interpreted as deboosted core prompt emission. We denote by $\tau = R_e/2\Gamma^2c$ the shell's angular timescale. Assuming the duration of the emission is negligible---i.e., instantaneous dissipation of energy in the shell---, the duration of the pulse as seen by an aligned observer is well approximated by the delay between the arrivals of photons from their beaming cone's axis and edge:
\begin{align}
\Delta t_{\rm on} & = \frac{R_e}{c} \left(1 - \cos 1/\Gamma \right) \\
& \sim \frac{R_e}{2 c \Gamma^2} \\
& = \tau
\label{eq:tau}
\end{align}

Assuming the dissipation is instantaneous simplifies much of the derivation below. We will adopt this hypothesis as this is a first exploration of this model and it allows to carry out the analytical development further. In practice, the relationship between $\tau$ and $\Delta t_{\rm on}$ depends on the prompt dissipation mechanism. Different mechanisms would affect the light curve profiles of the flares we will obtain below, but not the general features of the model. In Sec.~\ref{sec:grb}, we will discuss this point related to GRB emission mechanisms in more detail.

Therefore, writing $\Delta t_{\rm on} = \tau \sim (1 - \beta)R_e/c$ and using Eqs.~\ref{eq:1} and \ref{eq:2}, we arrive at this new form for the flare peak time:
\begin{equation}
t_{\rm flare} = \tau S + t_{\rm ej}
\label{eq:tf}
\end{equation}
where we used $\Delta t_{\rm on}=t_{\rm on}-t_{\rm ej}$.

Furthermore, the bolometric luminosity observed by the aligned observer is approximately $L_{\rm on} \sim E_{\rm iso}/\Delta t_{\rm on}$, where $E_{\rm iso}$ is the isotropic-equivalent source-frame dissipated energy in the shell \new{responsible for the emission of the pulse}. Using Eq.~\ref{eq:l}, we finally obtain \new{the luminosity of the corresponding flare as seen by a slightly off-axis observer}:
\begin{equation}
L_{\rm flare} = f_{\rm geo} S^{-3} \frac{E_{\rm iso}}{\tau}
\label{eq:lf}
\end{equation}

In our picture, the shell's ejection occurs during the central engine activity, which lasts for a duration $T_{\rm CE}$ depending on the smaller-scale physics around the central engine and in the accretion disk. As we do not have a firm constraint on this duration $T_{\rm CE}$, we shall consider it as a control parameter in the model fits we carry out in Sec.~\ref{sec:morpho}. Indeed, $T_{\rm CE}$ should not be much larger than the duration of GRB prompt phases, 
\new{often estimated by}
$T_{90}$. Similarly, the durations of pulses in GRBs are generally less than a few seconds \citep[e.g.,][]{2018ApJ...855..101H}. Therefore, the $\Delta t_{\rm on}$ (or $\tau$) we can consider are also constrained, to being less than a few seconds. Finally, for the prompt gamma-ray pulses to transform to X-ray flares, the stretch factor must typically be $\mathcal{S} \lesssim 100$, as mentioned in the numerical example above. As we shall soon see, keeping reasonable values for $E_{\rm iso}$ also implies that $\mathcal{S} \lesssim 100$, due to the flare flux suppression by $\mathcal{S}^{-3}$ (Eq.~\ref{eq:l}). Therefore, we conclude from Eq.~\ref{eq:tf} that the flares explained in our off-axis mechanism cannot appear much later than $t_{\rm flare} \sim 1000~{\rm s}$. As mentioned in Sec.~\ref{sec:intro}, there seems to be a dichotomy both in post-peak decay slopes and in the $\Delta F / F$ distribution between early ($t_{\rm flare} < 1000~{\rm s}$) and late ($t_{\rm flare} > 1000~{\rm s}$) flares \citep{2011A&A...526A..27B}. Early flares may thus be a distinct subclass of flares produced by a specific mechanism and the natural production of such arrival times by our model further motivates to pursue this picture. Moreover, these early flares are by far the most numerous in the observed population: in the source rest frame (i.e., redshift-corrected), $80\%$ of X-ray flares occur less than 260 s after the prompt trigger \citep{2016ApJS..224...20Y}; see however in Sec.~\ref{sec:classes} a discussion on the nature of these early flares and possible pollution by prompt emission.

\new{The typical values required for $\mathcal{S} \lesssim 100$ in our picture confirms that our flare model  applies only to slightly misaligned observers. Indeed, for large Lorentz factors this condition translates to
$\theta_j < \theta_v \lesssim \theta_j+\frac{10}{\Gamma}$. For $\Gamma\ge 100$, this is $\theta_v - \theta_j \le 10/\Gamma\le 0.1\lesssim \theta_j$.}

Inverting Eqs.~\ref{eq:tf} and Eqs.~\ref{eq:lf}, we conclude that, for observed flare properties $L_{\rm flare}$ and $t_{\rm flare}$ and assuming the shell's ejection time $t_{\rm ej}$, the isotropic-equivalent dissipated energy required to produce the flare is given as a function of $\tau$ by:
\begin{equation}
E_{\rm iso}(\tau) = \tau^{-2} \frac{L_{\rm flare}}{f_{\rm geo}} \left( t_{\rm flare} - t_{\rm ej}\right)^3
\label{eq:eiso}
\end{equation}

Assuming these $E_{\rm iso}$ and $\tau$ indeed reproduce the flare, the corresponding stretch factor is:
\begin{equation}
\mathcal{S} = \left( \frac{\tau L_{\rm flare}}{f_{\rm geo} E_{\rm iso}}\right)^{-1/3}
\end{equation}

In Fig.~\ref{fig:2dfit} (colored lines), we plot $E_{\rm iso}(\tau)$ assuming ejections times from $0$ to $T_{CE} = 80$ s, an estimate of GRB central engine activity duration we deduce from the average duration of \textit{Fermi}/GBM long bursts ($\bar{T_{90}} = 40$~s, \citealt{2020ApJ...893...46V}). Also, we assumed $f_{\rm geo} = 1/2$. In Fig.~\ref{fig:2dfit}, we chose the median source-frame (i.e., redshift-corrected) $t_{\rm flare}$ and $L_{\rm flare}$ in the \textit{Swift}/XRT afterglow flare sample: $\bar{t_{\rm flare}} = 85~{\rm s}$ and $\bar{L_{\rm flare}} = 1.8\times 10^{49}$~erg/s \citep{2010MNRAS.406.2113C,2016ApJS..224...20Y}. We also plotted iso-contours of the corresponding $\mathcal{S}$-factor. This figure allows us to outline which model parameter values are required to reproduce the typical X-ray flare and if these values are consistent with our physical setup and typical GRB quantities. According to the above discussion, the parameter space for our model is constrained by the following conditions:
\begin{itemize}
\item Shell ejection must occur during the primary central engine activity, thus $t_{\rm ej} \lesssim T_{CE}$;
\item We must maintain a slightly misaligned line of sight, thus $\mathcal{S} \gtrsim 10$;
\item Prompt gamma-rays must be deboosted to the X-ray band, thus $e_{\rm on}/e_{\rm off} = \mathcal{S} \lesssim 100$;
\item An aligned observer should detect a classical prompt pulse with duration typically between 0.1~s and 5~s: $0.1~{\rm s} \lesssim \Delta t_{\rm on} \sim \tau \lesssim 10~{\rm s}$;
\item The shell must dissipate an energy in line with what is observed for single pulses in typical GRBs. For entire GRBs (as opposed to single pulses within a GRB lightcurve), the dissipated energies are in the range $E^{\rm GRB}_{\rm iso} = 10^{53\pm1}~{\rm erg}$ (e.g., \citealt{2006MNRAS.372..233A}). Assuming a few pulses per GRB, we can estimate that each pulse dissipates an energy on the order of $E_{\rm iso} = 10^{52.3 \pm 1}~{\rm erg}$. We can thus consider this range for the shell's single dissipation episode.
\end{itemize}

The region in parameter space respecting all of these conditions is colored in purple in Fig.~\ref{fig:2dfit}: for any $(\tau, E_{\rm iso})$ pair in this zone, there corresponds an ejection time $t_{\rm ej}$ during the central engine activity and an $\mathcal{S}$-factor of a misaligned line of sight such that the flare appears at $\bar{t_{\rm flare}}$ with peak luminosity $\bar{L_{\rm flare}}$. This figure shows that there is open parameter space available to our model to explain typical flares as deboosted core prompt dissipation. Of course, there is much parameter space available for late ejection times ($t_{\rm ej} \sim T_{CE}$), such that the arrival time of the flare is mainly due to delayed activity. More interestingly, there is also space with larger $\tau$ and smaller ejection times ($t_{\rm ej} \le 60$~s), where the light travel time from the core to the misaligned observer plays a role in the flare arrival time.

One may use plots like Fig.~\ref{fig:2dfit} to find solutions to flares in actual GRB afterglows, adapting $t_{\rm flare}$ and $L_{\rm flare}$. For brighter or later flares, Eq.~\ref{eq:eiso} shows that larger $E_{\rm iso}$ are required, and the available parameter space will shrink. This equation also shows that allowing for larger ejection times eases the constraint on $E_{\rm iso}$. In Fig.~\ref{fig:2dfit} we bounded $t_{\rm ej}$ with a generic estimate for central engine activity duration, $t{\rm ej} \le T_{\rm CE} \lesssim 80~{\rm s}$. In fact, this bound should be on the duration of central engine activity \textit{in directions within the core}, $T^{\rm c}_{\rm CE}$. This duration may be different than the duration of the central engine activity on the misaligned observer's line of sight $T^{\rm LOS}_{\rm CE}$, measured by the actual GRB in which the flare we seek to explain appeared. Here, we will suppose that the central engine activity has the same duration in all directions and discuss this hypothesis in Sec.~\ref{sec:grb}.

For a given flare, one must also consider the actual temporal profile of the flare and seek solutions within this available parameter space that correctly fit the light curve. We will do so for two typical flare shapes in Sec.~\ref{sec:morpho}.

\section{Early flare visibility}
\label{sec:visi}
\subsection{Conditions for flare visibility}
\label{sec:cond}

\begin{table*}
\caption{Parameters for all flares with light curves represented in Fig.~\ref{fig:flares}. $t_{\rm ej}$: ejection time of the core shell responsible for the flare; $\tau^{\rm c}$: angular timescale $R_e/2\Gamma^2 c$ of the shell, where $\Gamma$ is the shell's Lorentz factor and $R_e$ is the dissipation radius; $t_{\rm flare}$: peak time of the flare; $w$: width of the flare, measured as the interval between the two times when the flux is $1/e$ times the peak flux; $L^{\rm flare} / L^{\rm ESD}$: ratio of the peak flux of the flare to the ESD flux at flare peak; $L^{\rm prompt}_{p, \rm BAT, on}$: peak pulse luminosity that an observer aligned with the shell would detect in the \textit{Swift}/BAT band. \new{All shells release the same isotropic equivalent energy $E^{\rm c}_{\rm iso} = 10^{53}~{\rm erg}$; all flares have $\mathcal{S} = 10$.} Note that our choice of gamma-ray band ({\it Swift}/BAT) and of emitted spectrum results in the approximate relation $L^{\rm prompt}_{p, \rm BAT, on} \sim 0.5 E^{\rm c}_{\rm iso} / \tau^{\rm c}$.}
\label{tab:flares}
\centering
\begin{tabular}{|l|ccc|cccc|}
\hline
\# & $t_{\rm ej}$ [s] & $\tau^{\rm c}$ [s] & $t_{\rm flare}$ [s] & $w/t_{\rm flare}$ & $L^{\rm flare} / L^{\rm ESD}$ & $L^{\rm prompt}_{p, \rm BAT, on}$ [erg/s]\\
\hline
\hline
A &  70 & 0.5 & 76 & 0.10 & 9.2 & $1.0 \times 10^{53}$ \\
B &  50 & 1 & 62 & 0.25 & 1.7 & $5.0\times 10^{52}$ \\
C &  80 & 1 & 92 & 0.17 & 9.7 & $5.0\times 10^{52}$ \\
D &  50 & 5 & 112 & 0.69 & 2.9 & $1.0\times 10^{52}$ \\
\hline
\end{tabular}
\end{table*}

The majority of X-ray flares occur early in the X-ray afterglow, during the early steep decay (ESD) phase. Within the context of a structured jet, we will now outline the conditions for the flares produced by our mechanism to appear above the ESD.

We suppose the observer lies at an angle $\theta_v$ from the jet axis. For this observer, the ESD will be produced by high-latitude emission from the last shell that flashed on their line of sight. Introducing the dissipated energy $E^{\rm LOS}_{\rm iso}$ of this last shell and its angular time scale $\tau^{\rm LOS}$, this ESD phase will have an approximate isotropic-equivalent bolometric luminosity of:
\begin{equation}
L^{\rm LOS}_{\rm ESD}(t) = \frac{E^{\rm LOS}_{\rm iso}}{\tau^{\rm LOS}} \left(\frac{t}{\tau^{\rm LOS}} \right)^{-3}
\end{equation}
for times larger than the end of the prompt emission.

Using the same notations as in Sec.~\ref{sec:outline} adding the superscript~$^{\rm c}$ for the shell in the core responsible for the flare, the flare's peak luminosity is still $L_{\rm flare} = f_{\rm geo} \mathcal{S}^{-3} E^{\rm c}_{\rm iso}/\tau^{\rm c}$ and the arrival time $t_{\rm flare} = \mathcal{S}\tau^{\rm c} + t_{\rm ej}$. We now introduce the jet structure, prescribing that, on average, the dissipated energies of shells in the core and on the $\theta_v$ line of sight are linked by\footnote{Note that the below definition pertains to the emitted energy in the prompt phase, in contrast to other definitions of jet structure which concern the initial kinetic energy in the jet.}:
\begin{equation}
\frac{E^{\rm LOS}_{\rm iso}}{E^{\rm c}_{\rm iso}} = \left(\frac{\theta_v}{\theta_j} \right)^{-a}
\label{eq:bo}
\end{equation}
where we chose a power-law for the energy structure for ease of the analytic development that will follow.

\new{The Lorentz factor of the material in the jet also possesses structure.} \new{We} adopt a similar power-law dependence for the material's average Lorentz factor:
\begin{equation}
\frac{\Gamma^{\rm LOS} - 1}{\Gamma_j - 1} = \left(\frac{\theta_v}{\theta_j} \right)^{-b}
\label{eq:ba}
\end{equation}
where $\Gamma_j$ is the average Lorentz factor of material in the jet core.

\new{The adopted value for $a$ and $b$ must be representative of the structure very close to the core jet ($\theta_j<\theta_v\lesssim 2\theta_j$), which can differ from  the shallow slopes $a, b \sim 5.5, 3.5$ inferred for the kinetic energy content of the jet in GW170817 \citep{GSPGY+2019} and that describe the structure up to very large angles $\theta_v \gtrsim 5 \theta_j$. Furthermore, these structure indices inferred from afterglow-data need not be equal to the indices we introduced in Eqs.~\ref{eq:bo} and \ref{eq:ba}, if the kinetic-to-prompt conversion efficiency is not constant within the jet.
As described in our plateau study \citep{BDDM2020}, steep slopes $a = 8$ and $b \gg 1$ are required to produce a  plateau behavior (their Fig. 1). 
Thus, we adopt in our numerical exploration  $a = 8$, $b = 6$ and  $\Gamma_j = 250$. 
This structure  of the Lorentz factor plays no role in the analytical developments of this section. However it is a prerequisite of our model that the misaligned observer's prompt emission is dominated by emission from the material on their line of sight. As shown in Appendix~\ref{sec:C}, this condition is fulfilled with these adopted values for the slightly misaligned regime of viewing angles we are considering ($\theta_j < \theta_v < 2 \theta_j$). 
}

\new{Of course, for both energy and Lorentz factor,} \new{the true structure may be more complicated and these prescriptions only concern the average material properties at a given angle from the jet axis.} \new{In particular, one should keep in mind that we suggest that flares are produced by individual shells in the core, for which properties such as Lorentz factor can deviate significantly from the average. A highly relativistic core with $\Gamma_j = 250$ that allows for ordinary prompt emission for an aligned observer does not exclude that some material in the core has a much lower}
\new{Lorentz factor, as preferred for the production of flares in our model (see Sec.~\ref{sec:outline}).}
\new{The fact that most GRB afterglows with flares only present one or two flares \citep{2010MNRAS.406.2113C}---in contrast with the many pulses in a GRB prompt phase---}\new{also} \new{suggests that only some of the shells produce flares.}

\new{With the assumed jet energy structure Eq.~\ref{eq:bo}, we obtain} the following flare--ESD contrast at the time of the flare:
\begin{align}
\frac{L_{\rm flare}}{L^{\rm LOS}_{\rm ESD}(t_{\rm flare})} & = f_{\rm geo} \mathcal{S}^{-3} \frac{\tau^{\rm c}}{\tau^{\rm LOS}} \left(\frac{\theta_v}{\theta_j} \right)^{a} \left( \frac{t_{\rm flare}}{\tau^{\rm c}}\right)^3 \\
& = f_{\rm geo} \frac{\tau^{\rm c}}{\tau^{\rm LOS}} \left(\frac{\theta_v}{\theta_j} \right)^{a} \left( \frac{\mathcal{S}^{-1}t_{\rm flare}}{\tau^{\rm c}}\right)^3 \\
& = f_{\rm geo} \left( \frac{\tau^{\rm c}}{\tau^{\rm LOS}} \right)^2 \left(\frac{\theta_v}{\theta_j} \right)^{a} \left(1 - \frac{t_{\rm ej}}{t_{\rm flare}} \right)^{-3}
\label{eq:contrast}
\end{align}
where we used $t_{\rm flare} = \mathcal{S}\tau^{\rm c} + t_{\rm ej}$, i.e., $\mathcal{S}^{-1} t_{\rm flare} = \tau^{\rm c} \left(1 - t_{\rm ej}/t_{\rm flare} \right)^{-1}$.

First, it appears from Eq.~\ref{eq:contrast} that a steeper structure favors the appearance of flares during the ESD. While this is true---it is simply that the ESD is dimmer for a steep structure---, one must bear in mind that a steeper structure also suppresses the misaligned prompt emission, and thus hinders the parent GRB detection and flare observation altogether. A finer calculation considering the likeliness of observing the GRB taking the structure into account is thus called for, see Sec.~\ref{sec:obs} for details.

Second, we find that, for fixed $\tau$'s, a later ejection time favors bright flares. This was already noted in our discussion of Fig.~\ref{fig:2dfit} and Eq.~\ref{eq:eiso}: one always has $t_{\rm flare} = \mathcal{S}\tau^{\rm c} + t_{\rm ej} \ge t_{\rm ej}$ and a larger $t_{\rm ej}$ favors a small $\mathcal{S}$ and thus a smaller suppression of flare flux by the $\mathcal{S}^{-3}$ factor. In the context of a flare during the ESD phase, it is clear that attributing a larger portion of the flare arrival time to ejection delay simply allows the flare to appear when the ESD flux is lower, as is clear in Eq.~\ref{eq:contrast}.

Finally, the appearance of the flare is dependent on the ratio of the core and line-of-sight \new{angular timescales} $\tau$. 
The $\tau$'s depend on the material's Lorentz factor and dissipation radii. In general, the ratio of $\tau$ between two different lines of sight is not known, and is strongly model dependent.

On the one hand, if generally $\tau^{\rm c} < \tau^{ \rm LOS}$, a bright flare can only be obtained if the ejection time is close to the time of the flare, as we just discussed. This may appear as a strong condition and leads us to discuss once more whether the central engine activity must be the same in all directions. If this activity is more prolonged in the core as compared to misaligned directions, there is more flexibility for large $t_{\rm ej}$ and bright flares, even if $\tau^{\rm c} < \tau^{ \rm LOS}$. On the other hand, if $\tau^{\rm c} > \tau^{ \rm LOS}$ is possible, this constraint does not apply and a flare can be produced without imposing a strict constraint on the duration of the central jet activity. Once again, we find that the central engine activity duration is a sensitive point in our model, and we will discuss more in Sec.~\ref{sec:grb}.

It should finally be noted that the above considerations are valid for the average pulse. Depending on the actual pulse shape and luminosity, Eq.~\ref{eq:contrast} may under- or over-estimate the actual contrast by a factor of a few. Also, spectral effects, not included in this section, could affect the flare visibility but are expected to remain moderate since both the ESD and the flares shine primarily in the X-rays.

\subsection{A natural mechanism for narrow flare production}
\label{sec:natural}

As mentioned in the introduction, a salient property of X-ray flares is their temporal aspect ratio $w / t_{\rm flare}$, where $w$ is the flare width. Flare widths have been measured in a number of ways: adopting for $w$ the width of a Gaussian fit to the flare light curve \citep[e.g.,][]{2007ApJ...671.1903C}; a smoothly broken power-law profile \citep{2016ApJS..224...20Y}; or using the \textit{Norris profile} \citep{2005ApJ...627..324N}, in which the width $w$ is naturally given by the time span between the two points before and after the peak when the flare flux is a factor of $e$ below the peak flux \citep{2010MNRAS.406.2113C,2011A&A...526A..27B}. Using this last definition, the aspect ratio $w/t_{\rm flare}$---which is independent of redshift---is found to be tightly distributed around $w/t_{\rm flare}$ = 0.23 with a standard deviation of 0.14 \citep{2010MNRAS.406.2113C}. In other words, flares' aspect ratios are small and nearly constant in the population. Any flare model must reproduce this fact.

We will now exhibit a natural mechanism built into our misaligned observer picture for flares by which the brighter flares tend to be thin. While we will later estimate $w/t_{\rm flare}$ for actual flare light curves in our model (Sec.~\ref{sec:hle}), we will temporarily focus on the temporal slope of flares in their post-peak phase, denoted by $\sigma = |\d \log L(t) / \d \log t|$. Analytically, we have a better grasp on $\sigma$ than on $w$, and for any given profile, larger $\sigma$ is related to smaller $w$, allowing us to focus on $\sigma$ for now.

We introduce the time-dependent stretch factor $S(t)$ (note the different font) which is simply the $\mathcal{S}$-factor of the core material the radiation of which is received at time $t$ by the misaligned observer. The calculation leading to Eq.~\ref{eq:l} is still valid when applied only to a strip of the shell from which the observer receives radiation at time $t$. The $f_{\rm geo}$ term changes with time, as the strip will progressively scan the geometry of the shell as seen from the observer's stand point. However, for a generic shell shape, $f_{\rm geo}$ will not change significantly, and we will thus have $L_{\rm flare}(t) \sim f_{\rm geo} L_{\rm on} S(t)^{-3}$. Finally, writing $S(t) = (t - t_{\rm ej})/\tau^{\rm c}$ as per Eq.~\ref{eq:tf}, we arrive at the following approximate flare time behavior:
\begin{equation}
L_{\rm flare}(t) = f_{\rm geo} L_{\rm on} \left( \frac{t - t_{\rm ej}}{\tau^{\rm c}} \right)^{-3}
\label{eq:s}
\end{equation}

Taking the logarithmic derivative of Eq.~\ref{eq:s} at $t = t_{\rm flare}$, we find that the initial post-peak decay index for the flare is:
\begin{align}
\sigma &= \left|-3 - \frac{3 t_{\rm ej}}{\mathcal{S}\tau^{\rm c}}\right|\\
&= \frac{3}{1 - \frac{t_{\rm ej}}{t_{\rm flare}}}
\label{eq:ss}
\end{align}

Therefore, coming back to Eq.~\ref{eq:contrast}, we find that the flare--ESD contrast is linked to the initial decay index by:
\begin{equation}
\frac{L_{\rm flare}}{L^{\rm LOS}_{\rm ESD}(t_{\rm flare})} = f_{\rm geo} \left( \frac{\tau^{\rm c}}{\tau^{\rm LOS}} \right)^2 \left(\frac{\theta_v}{\theta_j} \right)^{a} \left(\frac{\sigma}{3} \right)^{3}
\label{eq:sigma}
\end{equation}

This last equation shows that, under our interpretation of flares, and during the ESD, brighter flares---or simply those that rise above the continuum---tend to decay faster, and therefore be thinner, explaining the low values observed for $w/t_{\rm flare}$. This mechanism of selection of thin flares is particularly effective for small $\tau^{\rm c} / \tau^{\rm LOS}$. In this case, indeed, flares require even larger $\sigma$ (i.e., must be even thinner) to appear above the continuum. However, as mentioned in Sec.~\ref{sec:cond}, the relative values of $\tau^{\rm c}$ and $\tau^{\rm LOS}$ in GRB jets is uncertain.  

Moreover, the width of flares in our picture is $w \propto \mathcal{S} \Delta t_{\rm on} \sim \mathcal{S} \tau^{\rm c}$, because the prompt pulse duration  transforms like the photon arrival times to the off-axis line of sight. 
It therefore follows that the aspect ratio will be:
\begin{equation}
\frac{w}{t_{\rm flare}} 
= \eta\frac{\mathcal{S}\tau^{\rm c}}{\mathcal{S}\tau^{\rm c} + t_{\rm ej}}\, ,
\end{equation}
where $\eta$ is a proportionality constant depending on how exactly the width is measured.

\new{As long as the  arrival time is dominated by the angular effect
($\mathcal{S}\tau^{\rm c}$) and not by the ejection time, i.e. $t_\mathrm{ej}/\mathcal{S}\tau^{\rm c} \ll 1$, all flares should have the same aspect ratio, in agreement with the observed tight distribution. Dispersion is expected due to flares with late ejection times and smaller values of $\mathcal{S}$, i.e. to flares where $t_\mathrm{ej}/\mathcal{S}\tau^{\rm c} \lesssim 1$.}
The explanation of the small scatter in aspect ratio is thus tightly linked to the expected range in $t_{\rm ej}$ and $\mathcal{S}$, which we discuss in Sec.~\ref{sec:grb}.

\subsection{X-ray flares from prompt dissipation in the jet core}
\label{sec:hle}

For more concreteness we will now prescribe the emission physics for the core shell and study the production of flares for the misaligned observer. We \new{assume $\theta_{j}=0.1$ for the opening angle of the core jet and}
place the observer at $\theta_v = 0.13~{\rm rad}$, at the same location as in our previous computation of plateau emission (\citealt{BDDM2020}, Fig.~1). \new{The spherical shells responsible for the flares are in the core jet and, geometrically, cover the entire core of the jet up to a latitude of $\theta = \theta_j = 0.1$~rad.} We suppose they instantaneously radiate a source-frame isotropic equivalent energy of $E^{\rm c}_{\rm iso} = 10^{53}$~erg while moving at a Lorentz factor of \new{$\Gamma^\mathrm{c} = 100$} in the radial direction. This setup corresponds to $\mathcal{S} = 10$. We assign the shells with different values of $t_{\rm ej}$ and $\tau^{\rm c}$. We determine the light curves of the emission from these shells by integrating on equal-arrival-time strips on the shells corresponding to the angular exploration from the near to the far edge of the shell. All the light curves that we show are integrated in the \textit{Swift}/XRT band $\mathcal{B}_{\rm XRT} = [0.3-10]~{\rm keV}$ \citep{2004ApJ...611.1005G,2005SSRv..120..165B}. 

In order to produce the ESD radiation, we consider a shell along the observer's line of sight geometrically covering the entire jet from the line of sight to the core, which dissipates an energy $E^{\rm LOS}_{\rm iso} = 1.2\times 10^{52}$~erg, consistent with the $E_{\rm iso}^{\rm c}$ above and a \new{steep} power-law jet energy structure (Eq.~\ref{eq:bo}) with $a = 8$ \new{immediately outside the core as mentioned previously.} The shell producing the ESD is supposed to be the last flashing shell, to which we assign an ejection time of $t_{\rm ej}^{\rm LOS} = 30~{\rm s}$ and a pulse duration of $\tau^{\rm LOS} = 2~{\rm s}$. The line-of-sight shell has a Lorentz factor 
\new{$\Gamma^\mathrm{LOS}=50$, corresponding to a power-law structure (Eq.~\ref{eq:ba}) with a core average Lorentz factor $\Gamma_j = 250$ and a slope $b = 6$.} \new{We recall that this is the same jet structure as assumed in our previous work on plateaus \citep{BDDM2020}}.

For all core shells, we adopt a comoving emission spectrum of broken power-law shape, with low- and high-energy slopes $\alpha + 1= -0.1$ and $\beta  + 1= -1.2$ (average ``$F_\nu$'' slopes found in the GRB prompt phase, \citealt{2021arXiv210313528P}) and a shell-frame peak energy $E_p' = 1~{\rm keV}$, corresponding to the average observed $E_p$ of $\sim 200~{\rm keV}$ for an aligned observer and a Lorentz factor of 100. With their $\mathcal{S} = 10$, all the flares would thus have a peak energy of $20~{\rm keV}$.

\begin{figure}
\includegraphics[width=\linewidth]{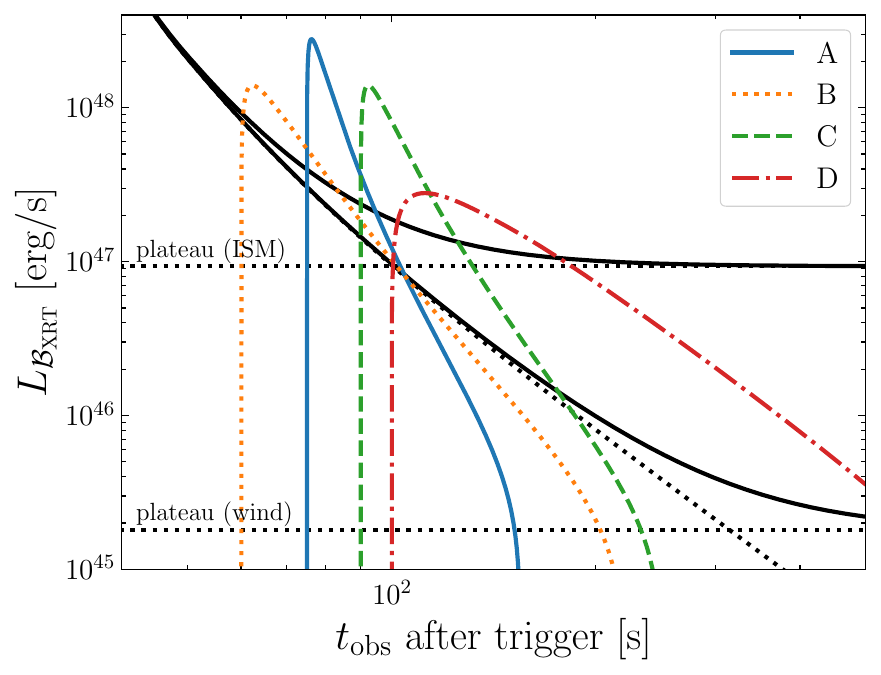}
\caption{X-ray luminosity expected from flashing core shells, viewed as flares by an off-core observer. Colored lines: luminosity from the core shells. Black lines: ESD signal, produced by material along the line of sight, along with typical plateau levels predicted by our misaligned observer plateau model. The parameters for the shells are reported in Tab.\ref{tab:flares}.}
\label{fig:flares}
\end{figure}

In Fig.~\ref{fig:flares}, one can find the resulting flare light curves for the core shells (colored lines) and the line of sight shell (ESD, black line). For completeness, we also added the level of a plateau predicted by our misaligned plateau model (\citealt{BDDM2020}, Eq.~10), with parameters exactly as in their Fig.~1, \new{with an external density $n = 1~{\rm cm}^{-3}$ for the uniform circum-burst medium hypothesis and a wind parameter $A_* = 0.1$ for the wind hypothesis}.
With these values adopted, it is as if the core shell dissipated $\eta_{\gamma} \sim 10\%$ of the initial available kinetic energy in the core \new{to produce a flare}, and the rest served to produce the afterglow plateau. \new{While this seems large at face value, it is consistent with typical prompt emission efficiencies estimated from data \citep{BNDP2015}}. Furthermore, the actual energy involved may be much less than this amount, if the flare producing shell is narrower than the jet's core (see discussion in Sec.~\ref{sec:morpho}).

The parameters for the flares can be found in Tab.~\ref{tab:flares}: we provide the $t_{\rm ej}$ and $\tau$ and report the flare time $t_{\rm flare}$, the aspect ratio $w/t_{\rm flare}$ as determined with the two points in time corresponding to a flux smaller by a factor of $e$ than the peak flux, the contrast $L_{\rm flare}/L_{\rm ESD}$ and, finally, the peak luminosity in the \textit{Swift}/BAT band ($[15-350]~{\rm keV}$, \citealt{2004ApJ...611.1005G,2005SSRv..120..143B}) that an observer would detect if they were aligned with the shell.

First, Fig.~\ref{fig:flares} shows that our model is capable of producing thin flares with occurrence times and luminosities consistent with typically observed flares. \new{We note that the aspect ratio of the flare in a more realistic approach would be slightly affected by the intrinsic duration of the dissipation process in the core jet, leading to a less steep rise, and by the angular size of the shell, as briefly discussed in Sec.~\ref{sec:morpho} where we fit our model to  \textit{Swift}/XRT data of two flares}. The small values of the aspect ratio are one of the most puzzling feature of observed flares. Therefore the low values of $w/t_\mathrm{flare}$ listed in Tab.~\ref{tab:flares} are especially encouraging. Second, Tab.~\ref{tab:flares} shows that, as seen on axis, these flares would produce peak luminosities of $\sim 10^{52}$~erg/s in the BAT band, therefore their properties are consistent with being able to reproduce the gamma-ray emission.

Furthermore, we find that, at given $\tau^{\rm c}$ (e.g., flares B and C), thinner flares tend to have a larger flare--ESD luminosity ratio, as discussed in Sec.~\ref{sec:natural}, though we find the effect to be more pronounced than expected. Similarly, for a fixed flare--ESD luminosity contrast (e.g., flares A and C), it appears that a larger $\tau^{\rm c}$ results in a flare with \new{a larger aspect ratio}, in line with Eq.~\ref{eq:sigma} (recall that a larger $\sigma$ is equivalent to a thinner flare). We also find that the larger $t_{\rm ej} /t_{\rm flare}$ (still $\le 1$ of course), the steeper the post-decay phase and the thinner the flare, as expected from Eq.~\ref{eq:ss}. The extreme case here being flare A, with $t_{\rm ej}/t_{\rm flare} \sim 92\%$ (because of its small $\tau^{\rm c}$) resulting in an extremely steep initial decay, with $\sigma \gtrsim 6$. \new{Finally, we note that obtaining later flares requires a larger $\tau^{\rm c}$ (e.g., flare D). This, however, produces wider and dimmer flares, as predicted in Sec.~\ref{sec:outline}.} In conclusion, this application of the model with instantaneous dissipation in the shells is consistent with the analytical results and observed trends in X-ray flares.

\section{Flare morphology}
\label{sec:morpho}

\begin{table*}
\centering
\caption{Best-fit source-frame parameters for XRT data of flares in GRB060719 and GRB100816A with our misaligned-observer interpretation of flares. We present parameters both constraining the ejection time to $t_{\rm ej} \le T_{90, \rm RF}$ or leaving it unconstrained. Notations for flare parameters are the same as in Tab.~\ref{tab:flares}.}
\label{tab:morpho}
\begin{tabular}{|l|ll|ll|}
\hline
GRB & \multicolumn{2}{c|}{060719 ($z = 1.532$, $T_{90, \rm RF} =26.4\pm4.5$)} & \multicolumn{2}{c|}{100816A ($z = 0.8034$, $T_{90, \rm RF} =1.6\pm0.3$)} \\
 & $t_{\rm ej}$ free & $t_{\rm ej} < T_{90, \rm RF}$ & $t_{\rm ej}$ free & $t_{\rm ej} < T_{90, \rm RF}$ \\
\hline
\hline
$E_{\rm iso}$ [erg]& $6.9\times10^{52}$ & $1.0\times10^{53}$ & $3.8\times10^{52}$ & $1.8\times10^{53}$ \\
$\Gamma$ & 119 & 87 & 78 & 90 \\
$\Delta \theta$ [rad] & 0.033 & 0.048 & 0.035 & 0.041 \\
$t_{\rm ej}$ [s] & 59 & 31 & 42 & 1.8 \\
$\tau$ [s] & 1.0 & 2.1 & 3.3 & 4.0 \\
$\mathcal{S}$ & 17 & 18  & 8.6 & 15 \\
\hline
\end{tabular}
\end{table*}

Beyond the width of flares, it appeared in the first catalog \citep{2007ApJ...671.1903C} that flares presented a variety of morphologies, with rising and decaying phases being fast or slow, i.e., exponential or power-law profiles. In terms of actual rise and decay times, virtually all flares decay in a longer time than they rise \citep{2010MNRAS.406.2113C}. However, the temporal profiles near flare peak define either very localized peaks in the case of fast rise and decay, or rounder flares in the case of slow rise and decay. Here, we will further assess the capabilities of our misaligned-observer interpretation of flares by studying which flare morphology it is able to capture.

\begin{figure*}
\centering
\resizebox{\hsize}{!}{\includegraphics{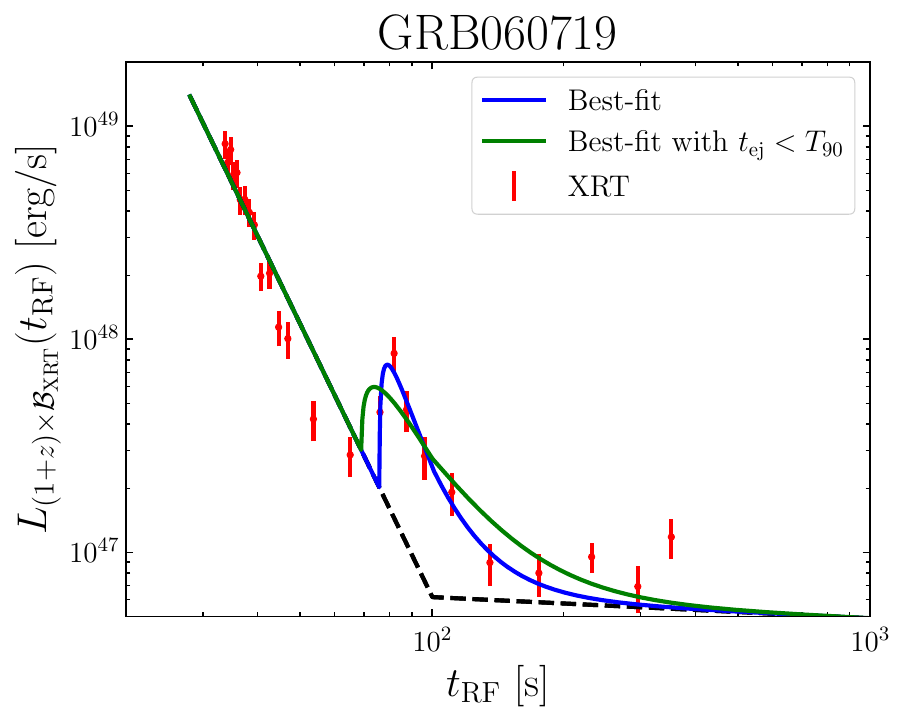}{\includegraphics{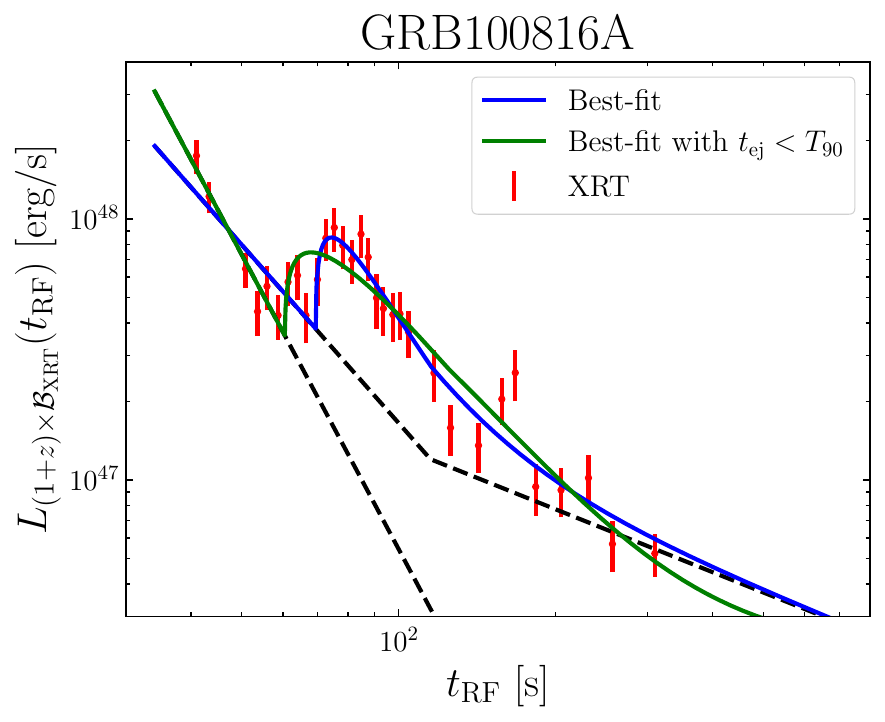}}}
\caption{Joint continuum and excess fitting to source-frame XRT data for two flares with different morphologies, using our misaligned core shell dissipation model. Best fits are shown both constraining the shell ejection time to $\le T_{90, \rm RF}$ (green) or not (blue). The dashed black line shows the best-fit continuum, which differs between the free-$t_{\rm ej}$ and constrained-$t_{\rm ej}$ fits only for GRB100816A (right). See Tab.~\ref{tab:morpho} for the corresponding best-fit parameters.}
\label{fig:morpho}
\end{figure*}

We seek to fit our model to GRB060719 (at $z = 1.532$) and GRB100816 (at $z = 0.8034$), which we choose because they feature flares peaking at $t_{\rm flare, \rm RF} \sim 90~{\rm s}$ in the source frame in both cases. The first presents a slow rise followed by a fast peak, while the second has a rounder peak and a slower decline, thus representing the variety in flare morphology. In Fig.~\ref{fig:morpho}, one can find the XRT data points for these two flares corrected for redshift. We applied the redshift correction \new{by determining the source-frame time with $t_{\rm RF} = t_{\rm obs} / (1 + z)$ and source-frame luminosity in the redshifted band with:}
\begin{equation}
L_{(1 + z) \times [\nu_{1, \rm obs}, \nu_{2, \rm obs}]} = \frac{1}{1 + z} 4\pi D_L(z)^2 F_{[\nu_{1, \rm obs}, \nu_{2, \rm obs}]}
\end{equation}
where $F_{[\nu_{1, \rm obs}, \nu_{2, \rm obs}]}$ is the observer-frame flux and, in our case, $[\nu_{1, \rm obs}, \nu_{2, \rm obs}] = \mathcal{B}_{\rm XRT} = [0.3-10]~{\rm keV}$. We determined the luminosity distance $D_L(z)$ using a generic flat world model with $H_0 = 70~{\rm km/s/Mpc}$, $\Omega_{\rm M} = 0.3$ and $\Omega_{\Lambda} = 0.7$. The observer-frame data were retrieved from the \textit{Swift}/XRT online archive \citep{2007A&A...469..379E,2009MNRAS.397.1177E}.

We proceed by fitting the source-frame XRT light curve by the sum of a broken power law \new{light curve} representing the continuum component and our flare model from core shells representing the excess \new{calculated as in Sec.~\ref{sec:hle}}. For a reminder, the off-axis shell model has the following parameters: isotropic-equivalent dissipated energy $E_{\rm iso}$, shell Lorentz factor $\Gamma$, angular distance between the observer and the shell's close edge $\Delta \theta = \theta_v - \theta_j$, shell ejection time $t_{\rm ej}$ and decay time scale $\tau$. The emitted spectrum adopted is the same as in Sec.~\ref{sec:hle}.

We seek the best-fit model as determined by minimizing a $\chi^2$ statistic under the same parameter constraints as in Sec.~\ref{sec:outline}: $E_{\rm iso}$ in the range $10^{52.3 \pm 1}$~erg, $\tau$ in the 0.1--5~s range, $\mathcal{S}$ must be $\gtrsim 10$. Concerning $t_{\rm ej}$, we consider two different conditions: Either we let it vary freely up to $t_{\rm flare, \rm RF}$ or we bound it by $T_{90, \rm RF}$, \new{the source-frame GRB duration}. Coming back to our discussion on central engine activity, the latter constrained-$t_{\rm ej}$ condition hypothesizes that this activity's duration is the same in all directions to the central engine, therefore having measured it through $T_{90}$ along the line of sight constrains it in the core; the former free-$t_{\rm ej}$ condition does not make this hypothesis. In Fig.~\ref{fig:morpho} we show the best fits under both these conditions; best-fit source-frame parameters can be found in Tab.~\ref{tab:morpho}.

First, it seems that satisfactory fits can be found to the light curve with reasonable parameter values: in both flares, the energies remain within the allowed region, $\Gamma \sim 100$ as anticipated in Sec.~\ref{sec:outline} and the $\mathcal{S}$-factors are $\gtrsim 10$, within the slightly misaligned regime. As a result, $t_{\rm ej}<0.5t_{\rm flare}$, meaning both the delayed ejection and the light travel time effects are at play.

\new{In these fits as in our examples of Sec.~\ref{sec:hle}, we supposed that the shells geometrically covered all the core, such that the real jet dissipated energy is:}
\begin{equation}
E_{\rm real}^{\rm c} \sim \frac{\theta_j^2}{2} E_{\rm iso}^{\rm c}
\end{equation}

\new{By changing the sizes of the shells (i.e., their angular diameter), we found that the observable section of the flare light curves do not change. The only signature of the size of the shell is the sharp cut off in flux produced by the far edge of the shell, and which can be noted for flares A, B, and C in Fig.~\ref{fig:flares}. In all cases however, this occurs at a flux level much lower than the underlying continuum and is therefore not visible to observers. In principle, the angular size of the shells are not constrained by the physics in the jet, apart from causality arguments implying their angular size cannot be smaller than $1/\Gamma$. As the shell size has little influence on the resulting flare, it allows some liberty to decrease the actual energy budget $E_{\rm real}^{\rm c}$ of the model.}

Second, it is clear that the hypothesis of instantaneous dissipation in the comoving frame leads to a sharp rise in the flares, and therefore allows our model to better fit fast-rising morphologies such as in GRB100816A rather than slow-rising ones like in GRB060719. 

\new{While instantaneous dissipation was practical for the analytical calculations above, it is not a requirement of our model. 
An intrinsic duration of dissipation will mainly affect the rise of the flare. However the properties of the declining phase---notably, the initial decay slope discussed in Sec.~\ref{sec:natural}---should remain}, as this phase is dominated by the angular exploration of the shell to latitudes further from the observer (this is equivalent to the situation in rise-decay ratios for prompt pulses, see Sec.~4.4 of \citealt{BG2016}). In particular, the transformation of the duration of a pulse to the duration of the flare (Eq.~\ref{eq:tf}) is still valid, when restricted to the declining phase of the pulse and flare.

Similarly, a shell shape different than circular or a unequally bright shell would change the profile of the rising phase. The circular shape we adopted is not particularly physically motivated, and the uniform shell brightness allowed to simplify some derivations; different prescriptions would not change the main features of our model. We thus conclude that an intrinsic dissipation duration and other shell shapes would allow to better capture slow-rising flares such as in GRB060719. However, the declining phase in both examples is well captured by the model, whether steep (GRB060719) or slow (GRB100816A).

Third, it is obvious that letting the ejection time run free makes for much better fits. This is true for GRB060719, where the $t_{\rm ej} \le T_{90, \rm RF}$ restriction seems to not allow the flare to peak at the right time. A similar issue, though less pronounced, occurs for GRB100816A. We did not statistically compare the goodness of fit of the free-$t_{\rm ej}$ and restricted-$t_{\rm ej}$ models. Nonetheless these two examples show that this hypothesis plays an important role in our model for flares, and we discuss it in more detail in Sec.~\ref{sec:grb}.

\section{Discussion}
\label{sec:discussion}

\subsection{Summary}
Motivated by modeling for plateaus in GRB afterglows in the physical setting of a slightly misaligned observer to a structured jet \citep{EG2006,BDDM2020}, we have presented a new model to interpret flares in the same setup. We suggest that flares can be produced by prompt dissipation in the jet's core that appears in the X-ray band rather than the gamma-rays because the core is less boosted to a misaligned observer than to a on-axis observer. In our picture, the delay in the flare observation with respect to the prompt emission is a combination of a purely geometrical effect, linked to the photon travel time from the core to the misaligned line of sight, and an intrinsic effect linked to the finite duration of the central engine activity, resulting in delays in ejection of different shells.

Writing down the transformations of photon arrival time, spectrum and luminosity from an aligned to a misaligned line of sight, we found that the typical X-ray flare could indeed be explained as deboosted core jet prompt emission (Sec.~\ref{sec:outline}). In doing so, we outlined the typical properties required for the core shells responsible for the flares and the expected properties for the resulting flares: those shells have rather low Lorentz factors $\Gamma \sim 100$, carry energies typically on the higher end of those dissipated in single pulses of GRBs $E_{\rm iso} \gtrsim 10^{52}$~erg. The flares thus produced naturally occur early in the afterglow ($t_{\rm flare} \le 1000~{\rm s}$), when the majority of flares are observed in XRT light curves.

In the early afterglow phase, the continuum is dominated by the ESD. We therefore analytically studied the conditions for appearance of flares during the ESD, assuming it was produced by high-latitude emission from the last flashing shell on the misaligned observer's line of sight (Sec.~\ref{sec:visi}). We found that flare visibility is favored by late shell ejection times $t_{\rm ej}$ and long shell decay timescales $\tau_{\rm c}$. We also demonstrated a mechanism present in our model by which brighter flares---or simply those that are able to appear above the continuum---tend to be narrower, echoing the observation that most flares have aspect ratios $w/t_{\rm flare} \lesssim 0.5$. Furthermore, synthetic light curves confirmed these trends between brightness and aspect ratio, the prominent role of the ejection time in flare visibility and, overall, proved that typical flare widths are reproduced by our model (Sec.~\ref{sec:hle}).

Finally, we made fits of our model to two actual flares observed in XRT afterglows (Sec.~\ref{sec:morpho}). We chose these two examples so as to represent two different morphologies found in flares: slow-rise-fast-decay (GRB060719) and fast-rise-slow-decay (GRB100816A). We found satisfactory fits with reasonable parameter values and a setup within the slightly misaligned regime, as measured by $\mathcal{S} \sim 1  + |\Delta \theta \Gamma|^2 \gtrsim 10$. The model would better capture the slow rises assuming an intrinsic duration of the shell dissipation---our analytical work considered it instantaneous for simplicity.

These fits confirmed the role of the ejection time in defining the flare arrival time and the trade-off between shell energy and ejection time anticipated in Sec.~\ref{sec:outline}. Indeed, flare arrival time increases with both $\mathcal{S}$ (more misalignment) and $t_{\rm ej}$ (later ejection), flare luminosity however drastically decreases with $\mathcal{S}$. Thus, for a given flare, increasing $t_{\rm ej}$ allows to somewhat decrease $\mathcal{S}$ and the shell energy.

\subsection{Admissible prompt dissipation mechanisms}
\label{sec:grb}

In principle, our explanation of flares and our analytical results are compatible with any prompt dissipation mechanism, as long as the declining phases of gamma-ray pulses are due to the angular exploration of the shell, i.e., high-latitude emission. This will be the case for prompt mechanisms with negligible dissipation duration compared to the angular timescale, such as internal shocks \citep{1998MNRAS.296..275D,2000A&A...358.1157D} or macroscopic magnetic reconnection events at large distances from the central engine \citep{LB2003,KN2009,KC2015,BG2016,BLG2016,BBG2018}. It  could also be the case for models with smaller dissipation radii, such as dissipative photospheric models \citep[e.g.,][]{2005ApJ...628..847R,2017SSRv..207...87B}, if the central engine turns off sufficiently rapidly to mimic a steep decline as in high-latitude emission. All the other prompt-related parameters of the model such as energy and shell Lorentz factor are generic. Because of this last point, the occurrence of the  mechanism we propose is inevitable: any misaligned observer to a GRB jet is exposed to detecting delayed and deboosted prompt emission from the jet core. Our fit to the flare in GRB100816A occurring 100 s after trigger with $t_{\rm ej} \sim 0$ (Fig.~\ref{fig:morpho}, right) shows that even a null ejection delay can trigger the mechanism. There must therefore be some flares produced in this way. As already mentioned, these should be early flares because of the allowed parameter ranges. The statistical properties of the flares produced by our mechanism will be further discussed in Sec.~\ref{sec:obs} below.

One subtle point of our model is the admissible range for the ejection time of the core shells. As we mentioned, larger $t_{\rm ej}$ allows for brighter, thinner flares. \new{In our picture,} the shell ejection must naturally occur during the \new{first and only episode of} central engine activity. Having observed a GRB as a misaligned observer informs us about the central engine activity duration along one's line of sight which dominates the observed signal. This duration has no reason to be the same in the jet core where we posit the flare-producing shells lie. The example of GRB100816A (Fig.~\ref{fig:morpho}, right) shows that a better fit is found by allowing $t_{\rm ej}$ to reach typical GRB $T_{90}$ durations of $40~{\rm s}$ which, however, are much longer than the parent GRB's source-frame $T_{90}$ of 1.6~s. GRB060719 (Fig.~\ref{fig:morpho}, left) provides a less drastic example.

These results suggest the question of whether the central engine activity can be shorter or longer lasting depends on the ejection direction. Such variation in the activity duration around the central engine is prescribed upstream, by the physical conditions near the compact source. Because of, e.g., the significant interaction of the incipient jet with material near the compact object, it could be that the time interval during which relativistic material effectively emerges from the system depends on the latitude. One could speculate that the jet breaks out early and continues to eject matter through the core while the \new{off-axis} material experiences more interaction with the cocoon and thus relativistic material emergence is shorter-lived. To our knowledge, the comparison of relativistic ejection of different lines of sight remain to be studied in numerical simulations, and more fits of our model to XRT data must be done to determine whether the requirement of $t_{\rm ej} \ge T_{90}$ is a general feature.

Generally, a constant $w/t_{\rm flare}$ cannot be a natural consequence of a \textit{composite model}, in which $w$ and $t_{\rm flare}$ are set by unrelated causes. For example, in genuine late central engine activity \citep[e.g.,][]{2005ApJ...630L.113K,2006ApJ...636L..29P,2006Sci...311.1127D}, $w$ and $t_{\rm flare}$ are determined respectively by the duration and onset time of the second engine episode. In this case, the tuning of $w/t_{\rm flare}$ to a same value of the aspect ratio from one system to another is not clear, especially seeing the diversity of durations in the first episode, as suggested by GRB durations. \new{In models with changes in the reverse shock propagation medium \citep[e.g.,][]{2017MNRAS.472L..94H}, $w$ and $t_{\rm flare}$ are determined by the size and the propagation time up to the external medium accident, tuning these also seems unnatural. A similar limitation affects pictures for flares including a stratified propagation medium for the reverse shock \citep{2018MNRAS.474.2813L,2020MNRAS.495.2979A} or those with a delayed prompt dissipation \citep{2006A&A...455L...5G,2009ApJ...692..133Y,2015ApJ...803...10T}.} \new{We also note that such models also fall short of explaining the steep rises of some flares (e.g., in GRB060719, Fig.~\ref{fig:morpho}, left).}

Conversely, a constant $w/t_{\rm flare}$ is a natural consequence of models \new{with a single episode of central engine activity and a single episode of prompt dissipation}, such as ours and the model of \cite{BK2016}. In these cases, the aspect ratio is determined by a single transformation of prompt emission, in our case by geometrical effects, i.e., photon travel time. The fact that the distribution of rise-to-decay time ratios for flares closely follows that of GRB prompt pulses \citep{2010MNRAS.406.2113C} further encourages such models. In our picture, we showed in Sec.~\ref{sec:natural} that a small scatter in flare aspect ratio is obtained by diversity in $t_{\rm ej}$ and $\mathcal{S}$: we expect the geometrical effect ($\mathcal{S}\tau^{\rm c}$) to be dominant in shaping flare aspect ratio, thus producing a tight distribution, and the subdominant $t_{\rm ej}$ to introduce a small scatter. 

Another remarkable property of our model is that the flares naturally appear in the X-rays, even if gamma-rays are produced for an aligned observer. It is not obvious why the central engine should shine in the X-rays and with an increasing timescale of variability in other models. In fact, because of the misaligned nature of the observer in our picture, no simultaneous higher-energy counterpart (e.g., gamma-rays) is expected. Indeed, it seems that any super-gamma-ray photons which would eventually appear as gamma-rays to the misaligned observer should be suppressed by optical depth to pair production, scattering on pairs and on electrons \citep{MNP2019,MNP2019b}; the X-rays however are not affected. A thorough establishment of this prediction in our geometrical setup remains to be done. We checked that no simultaneous BAT counterpart was present in the two examples treated above in Sec.~\ref{sec:morpho}.  \new{While we have mainly discussed flares in the X-ray band, they are also present in the optical, where they are found to be statistically similar to X-ray flares \citep{2012ApJ...758...27L,2017ApJ...844...79Y} and to also occur mostly in the early afterglow \citep{2013ApJ...774....2S}. This could indicate a common origin for optical and X-ray flares as well, as suggested here. There remains to explore the optical signatures of our picture for flares. Surely the fitting of multi-wavelength data will further constrain the parameter space.}

\subsection{Observed and expected model consequences}
\label{sec:obs}

There are rather strong conditions for successful flaring in our model. \neww{Mainly, these are large $E_{\rm iso}$ and small $\Gamma$ in the shells responsible for the flares, as dictated by the transformation of luminosity from the aligned to misaligned lines of sight (Eq.~\ref{eq:lf}). These conditions are further restricted by finer effects, for example the fact that later GRB prompt pulses tend to be less energetic than earlier ones \citep{2010A&A...511A..43G}, whereas our model favors the appearance of flares from shells ejected later rather than earlier. Therefore,} we do not expect all the shells in a given GRB jet to produce visible flares, and we do not expect many flares in each GRB. Adding the condition of slightly misaligned line of sight to the GRB jet, we do not expect many GRBs with flares in general; the observation is that about one third of \textit{Swift} GRBs exhibit noticeable flares \citep{2016ApJS..224...20Y}. Discussing such statistics requires to consider many parameters: the allowed range for $t_{\rm ej}$; the jet structure, which conditions both the GRB trigger for off-axis observers and the flare-to-ESD contrast (Eq.~\ref{eq:contrast}); the Lorentz factors of the shells within the core. Such a statistical study would further root our model into flare observations. Moreover, as the physical setup is the same for our two models, statistics of both flaring and plateau behavior in GRB afterglow would shed further light on our models. \new{Such joint flare--plateau activity has already been noted \citep[e.g., GRB080129,][]{2009ApJ...693.1912G,2009ApJ...697.1044G} and a recent study has quantified the proportion of GRBs with plateaus and flares to $\sim 50\%$ among GRBs with flares \citep{2021arXiv211101041Y}. In this sub-sample, strong correlations are found between the energy dissipated in the parent GRB prompt phases and in the flares. This observation is encouraging for our picture of plateaus and flares, and we leave a thorough statistical study to future work.}

An interesting consequence of our model is the natural dichotomy between early and late flares, \new{with the former class being statistically dominant \citep{2014ApJ...788...30S}}. With $t_{\rm ej}$ limited to $\lesssim 100~{\rm s}$ (i.e., typical central engine activity), flares with $t_{\rm flare} \ge 1000~{\rm s}$ are only possible in our picture with large $\mathcal{S}$. However with $L_{\rm flare} \propto \mathcal{S}^{-3} E_{\rm iso}$, we do not expect these flares to be visible. This fact remarkably echoes the observation of a dichotomy in temporal behavior and $\Delta F / F$ distributions between early and late flares \citep{2011A&A...526A..27B}: a different origin for late flares is therefore reasonable.

It is also interesting that, for these early flares, \citet{2011A&A...526A..27B} found the relation $L_{\rm flare} \propto t_{\rm flare}^{-2.7 \pm 0.1}$ between flare arrival times and peak fluxes. Considering that $L_{\rm flare} \sim \mathcal{S}^{-3} L_{\rm on}$ and $t_{\rm flare} = \mathcal{S}\tau^{\rm c} + t_{\rm ej}$, the correlation spanned by varying $\mathcal{S}$ for constant $\tau$ and shell energies should be $L_{\rm flare} \propto t_{\rm flare}^{-3}$ exactly, in the absence of shell ejection delays. These delays allow however for later flares with the same luminosity, such that the slope is in fact slightly shallower than --3, as found. \new{Outlying flares of such luminosity-peak time correlations that are both late and bright (such as the ``giant flare'' GRB050502B, \citealt{2006ApJ...641.1010F}, or the late flares in GRB121027A, \citealt{2013arXiv1302.4876P}) are however out of reach of our model's parameter space.} Finally, for GRBs with many flares, we expect the later ones to have larger $\mathcal{S}$'s generally, considering the other parameters fixed. We thus expect them to be dimmer and softer; these trends are indeed found in the few GRBs with more than one flare \citep{2010MNRAS.406.2113C}.

If our picture for flares is correct, there are further consequences that we could check in the population. First, while the question of dependence of $T_{\rm CE}$ on the line of sight is delicate, we should generally expect that, if $T_{90}$ is large in a given GRB, then it should also be large on other lines of sight to the same GRB. As large ejection times favor flares, we expect long $T_{90}$ to be correlated with flaring activity. \new{A preliminary inspection of the \textit{Swift} archive seems to confirm this trend, though a thorough statistical study is called for.} Second, we describe early flares and plateaus as the consequence of misaligned lines of sight, thus we expect the jet structure to introduce a trend between flaring activity and dim prompt emission. Third, if there is a qualitative or quantitative difference in the prompt dissipation mechanism between the core and the lateral structure, it should appear when comparing prompt emission of GRBs with and without flaring or plateau activity. Indeed in our picture we expect the prompt emitting regions in these two cases to be located in different parts of the jet structure \new{in which, at least quantitatively, the physical conditions and thus prompt dissipation mechanism could be different}. With this last point, we understand that flares and plateaus could provide key insight into GRB physics, the structure of GRB jets and the difference between long and short GRBs in this respect.

Generally, while the present publication is only a first proposal and limited exploration of this mechanism for flares, the many consequences of this model on the population of flares that we have just listed makes it testable and provides motivation to deepen the statistical study of GRB afterglow flares. \neww{In confronting our model's predictions with flare statistics and possibly seeking to constrain the jet structure, we note three main points of caution. First, we expect some---yet unknown---degree of diversity in GRB jet structures and core jet opening angle. Second, one must carefully handle selection effects in flare samples linked to incomplete redshift data or X-ray coverage of the early afterglow of GRBs. Third, one must account for detection biases intrinsic to our model: in our picture, the detection of flares is subject to the detection of the prompt emission from the off-axis line-of-sight material, which is itself linked to the energy and Lorentz-factor structures of the jet.}

\subsection{On the nature of flares and the flare sample}
\label{sec:classes}

Our model posits a common origin for prompt emission and early X-ray flares, and is supported by some similarities between these phenomena. Consequently, it poses the question of the definition of flares and sample contamination in X-ray flares. All X-ray flare samples we mentioned in this paper \citep{2007ApJ...671.1921F,2007ApJ...671.1903C,2010MNRAS.406.2113C,2011MNRAS.410.1064M,2011A&A...526A..27B,2014ApJ...788...30S,2016ApJS..224...20Y} are selected only after visual inspection of the XRT light curves, in search for excess flux over a continuum. While this is justified for late flares (such as the late flare sample of \citealt{2011A&A...526A..27B}), the sample of early flares thus selected must be contaminated by prompt emission as well. Generally, a more physical definition of flares is warranted to better isolate this activity and thus define what exactly models should seek to reproduce. For example, in our picture, it seems that flares should be characterized by an absence of a counterpart in the higher-energy bands such as the BAT and a significantly \new{larger} hardness ratio than the underlying continuum emission. This definition excludes, for example, the sample collected in \citet{2014ApJ...795..155P}, the flares in GRB050820A \citep[][$t_{\rm flare} \sim 400~{\rm s}$]{2006ApJ...652..490C} and GRB110801A \citep[][$t_{\rm flare} \sim 200~{\rm s}$]{2011GCN.12228....1D} and the first two of the five flares in GRB060714 \citep[][$t_{\rm flare} = 80-100~{\rm s}$]{2007ApJ...665..554K}, which show a simultaneous BAT excess.

\neww{Furthermore}, we note that positing a common origin for GRB prompt emission and X-ray flares further adds to the discussion of the possibility that long GRBs, low-luminosity GRBs, X-ray flashes, etc. could be the manifestations of the same system viewed from different orientations \citep[e.g.,][]{2005ApJ...630.1003G,2005ApJ...625L..91R,GR2010}. That fact that the low-luminosity GRB031203 and GRB171205A present some phases of flat X-ray flux (respectively, \citealt{2005ApJ...625L..91R,2018A&A...619A..66D}) supports the idea that they could be misaligned events, in light of our plateau model. In light of the present flare model, the existence of X-ray flashes with contemporaneous X-ray and optical flares (such as XRF071031, \citealt{2009ApJ...697..758K}) also supports this view. The investigation of X-ray flashes as misaligned core prompt dissipation with a model as presented here could shed more light on the nature of these events. In any case, both a statistical study and light curve fits should be carried out.

\neww{Finally, we note that the present article and other recent afterglow modeling endeavors \citep[e.g.,][]{2020ApJ...893...88O,2020A&A...641A..61A,BGG2020} illustrate the strong effects of observer--jet geometry in shaping GRB prompt and afterglow emission, even without specifying the underlying dissipation mechanisms at play. The combination of these geometrical effects with structured jets makes for fruitful modeling grounds and allow for fresh views on the GRB phenomenon.}

\section{Conclusion}
\label{sec:conclusion}

In the same physical setup as a model to explain the plateau features in GRB afterglows, we exhibited a novel interpretation for X-ray flares in GRB afterglows. It relies on slightly misaligned lines of sight to a structured relativistic jet, in which the core's prompt dissipation is deboosted to the X-ray band for the off-axis observer and appears during the afterglow---typically, during the early steep decay---because of both the light travel time from the core to the observer and the intrinsic duration of the central engine ejection activity. From order-of-magnitude considerations to actual fits to \textit{Swift}/XRT data, we showed that this model is capable of explaining typical flares. Further, we showed how this model favors flares with small and tightly distributed aspect ratios, a salient property of X-ray flares. Overall, though a thorough statistical study is called for, it appears that there are many trends found in GRB afterglows flares that our picture naturally produces.

\section*{Acknowledgments}

We thank T. Matsumoto, R. Sari and O. S. Salafia for useful discussion. RD is supported by the ERC Advanced Grant ``JETSET: Launching, propagation and emission of relativistic jets from binary mergers and across mass scales'' (Grant No. 884631). PB was supported by the Gordon and Betty Moore Foundation, Grant GBMF5076 and by grant (no. 2020747) from the United States-Israel Binational Science Foundation (BSF), Jerusalem, Israel.

\section*{Data availability}

The data and software underlying this article will be shared on reasonable request to the corresponding author.


\bibliographystyle{mnras}
\bibliography{main}

\begin{thebibliography}{}
\makeatletter
\relax
\def\mn@urlcharsother{\let\do\@makeother \do\$\do\&\do\#\do\^\do\_\do\%\do\~}
\def\mn@doi{\begingroup\mn@urlcharsother \@ifnextchar [ {\mn@doi@}
  {\mn@doi@[]}}
\def\mn@doi@[#1]#2{\def\@tempa{#1}\ifx\@tempa\@empty \href
  {http://dx.doi.org/#2} {doi:#2}\else \href {http://dx.doi.org/#2} {#1}\fi
  \endgroup}
\def\mn@eprint#1#2{\mn@eprint@#1:#2::\@nil}
\def\mn@eprint@arXiv#1{\href {http://arxiv.org/abs/#1} {{\tt arXiv:#1}}}
\def\mn@eprint@dblp#1{\href {http://dblp.uni-trier.de/rec/bibtex/#1.xml}
  {dblp:#1}}
\def\mn@eprint@#1:#2:#3:#4\@nil{\def\@tempa {#1}\def\@tempb {#2}\def\@tempc
  {#3}\ifx \@tempc \@empty \let \@tempc \@tempb \let \@tempb \@tempa \fi \ifx
  \@tempb \@empty \def\@tempb {arXiv}\fi \@ifundefined
  {mn@eprint@\@tempb}{\@tempb:\@tempc}{\expandafter \expandafter \csname
  mn@eprint@\@tempb\endcsname \expandafter{\@tempc}}}

\bibitem[\protect\citeauthoryear{{Amati}}{{Amati}}{2006}]{2006MNRAS.372..233A}
{Amati} L.,  2006, \mn@doi [\mnras] {10.1111/j.1365-2966.2006.10840.x}, \href
  {https://ui.adsabs.harvard.edu/abs/2006MNRAS.372..233A} {372, 233}

\bibitem[\protect\citeauthoryear{{Ascenzi}, {Oganesyan}, {Salafia},
  {Branchesi}, {Ghirlanda}  \& {Dall'Osso}}{{Ascenzi}
  et~al.}{2020}]{2020A&A...641A..61A}
{Ascenzi} S.,  {Oganesyan} G.,  {Salafia} O.~S.,  {Branchesi} M.,  {Ghirlanda}
  G.,   {Dall'Osso} S.,  2020, \mn@doi [\aap] {10.1051/0004-6361/202038265},
  \href {https://ui.adsabs.harvard.edu/abs/2020A&A...641A..61A} {641, A61}

\bibitem[\protect\citeauthoryear{{Ayache}, {van Eerten}  \& {Daigne}}{{Ayache}
  et~al.}{2020}]{2020MNRAS.495.2979A}
{Ayache} E.~H.,  {van Eerten} H.~J.,   {Daigne} F.,  2020, \mn@doi [\mnras]
  {10.1093/mnras/staa1397}, \href
  {https://ui.adsabs.harvard.edu/abs/2020MNRAS.495.2979A} {495, 2979}

\bibitem[\protect\citeauthoryear{{Barniol Duran}, {Leng}  \&
  {Giannios}}{{Barniol Duran} et~al.}{2016}]{BLG2016}
{Barniol Duran} R.,  {Leng} M.,   {Giannios} D.,  2016, \mn@doi [\mnras]
  {10.1093/mnrasl/slv140}, \href
  {https://ui.adsabs.harvard.edu/abs/2016MNRAS.455L...6B} {455, L6}

\bibitem[\protect\citeauthoryear{{Barthelmy} et~al.,}{{Barthelmy}
  et~al.}{2005}]{2005SSRv..120..143B}
{Barthelmy} S.~D.,  et~al., 2005, \mn@doi [\ssr] {10.1007/s11214-005-5096-3},
  \href {https://ui.adsabs.harvard.edu/abs/2005SSRv..120..143B} {120, 143}

\bibitem[\protect\citeauthoryear{{Beloborodov} \&
  {M{\'e}sz{\'a}ros}}{{Beloborodov} \&
  {M{\'e}sz{\'a}ros}}{2017}]{2017SSRv..207...87B}
{Beloborodov} A.~M.,  {M{\'e}sz{\'a}ros} P.,  2017, \mn@doi [\ssr]
  {10.1007/s11214-017-0348-6}, \href
  {https://ui.adsabs.harvard.edu/abs/2017SSRv..207...87B} {207, 87}

\bibitem[\protect\citeauthoryear{{Beniamini} \& {Granot}}{{Beniamini} \&
  {Granot}}{2016}]{BG2016}
{Beniamini} P.,  {Granot} J.,  2016, \mn@doi [\mnras] {10.1093/mnras/stw895},
  \href {https://ui.adsabs.harvard.edu/abs/2016MNRAS.459.3635B} {459, 3635}

\bibitem[\protect\citeauthoryear{{Beniamini} \& {Kumar}}{{Beniamini} \&
  {Kumar}}{2016}]{BK2016}
{Beniamini} P.,  {Kumar} P.,  2016, \mn@doi [\mnras] {10.1093/mnrasl/slw003},
  \href {https://ui.adsabs.harvard.edu/abs/2016MNRAS.457L.108B} {457, L108}

\bibitem[\protect\citeauthoryear{{Beniamini} \& {Nakar}}{{Beniamini} \&
  {Nakar}}{2019}]{BN2019}
{Beniamini} P.,  {Nakar} E.,  2019, \mn@doi [\mnras] {10.1093/mnras/sty3110},
  \href {http://cdsads.u-strasbg.fr/abs/2019MNRAS.482.5430B} {482, 5430}

\bibitem[\protect\citeauthoryear{{Beniamini}, {Nava}, {Duran}  \&
  {Piran}}{{Beniamini} et~al.}{2015}]{BNDP2015}
{Beniamini} P.,  {Nava} L.,  {Duran} R.~B.,   {Piran} T.,  2015, \mn@doi
  [\mnras] {10.1093/mnras/stv2033}, \href
  {https://ui.adsabs.harvard.edu/abs/2015MNRAS.454.1073B} {454, 1073}

\bibitem[\protect\citeauthoryear{{Beniamini}, {Barniol Duran}  \&
  {Giannios}}{{Beniamini} et~al.}{2018}]{BBG2018}
{Beniamini} P.,  {Barniol Duran} R.,   {Giannios} D.,  2018, \mn@doi [\mnras]
  {10.1093/mnras/sty340}, \href
  {https://ui.adsabs.harvard.edu/abs/2018MNRAS.476.1785B} {476, 1785}

\bibitem[\protect\citeauthoryear{{Beniamini}, {Duque}, {Daigne}  \&
  {Mochkovitch}}{{Beniamini} et~al.}{2020a}]{BDDM2020}
{Beniamini} P.,  {Duque} R.,  {Daigne} F.,   {Mochkovitch} R.,  2020a, \mn@doi
  [\mnras] {10.1093/mnras/staa070}, \href
  {https://ui.adsabs.harvard.edu/abs/2020MNRAS.492.2847B} {492, 2847}

\bibitem[\protect\citeauthoryear{{Beniamini}, {Granot}  \& {Gill}}{{Beniamini}
  et~al.}{2020b}]{BGG2020}
{Beniamini} P.,  {Granot} J.,   {Gill} R.,  2020b, \mn@doi [\mnras]
  {10.1093/mnras/staa538}, \href
  {https://ui.adsabs.harvard.edu/abs/2020MNRAS.493.3521B} {493, 3521}

\bibitem[\protect\citeauthoryear{{Bernardini}, {Margutti}, {Chincarini},
  {Guidorzi}  \& {Mao}}{{Bernardini} et~al.}{2011}]{2011A&A...526A..27B}
{Bernardini} M.~G.,  {Margutti} R.,  {Chincarini} G.,  {Guidorzi} C.,   {Mao}
  J.,  2011, \mn@doi [\aap] {10.1051/0004-6361/201015703}, \href
  {https://ui.adsabs.harvard.edu/abs/2011A&A...526A..27B} {526, A27}

\bibitem[\protect\citeauthoryear{{Burrows} et~al.,}{{Burrows}
  et~al.}{2005a}]{2005SSRv..120..165B}
{Burrows} D.~N.,  et~al., 2005a, \mn@doi [\ssr] {10.1007/s11214-005-5097-2},
  \href {https://ui.adsabs.harvard.edu/abs/2005SSRv..120..165B} {120, 165}

\bibitem[\protect\citeauthoryear{{Burrows} et~al.,}{{Burrows}
  et~al.}{2005b}]{2005Sci...309.1833B}
{Burrows} D.~N.,  et~al., 2005b, \mn@doi [Science] {10.1126/science.1116168},
  \href {https://ui.adsabs.harvard.edu/abs/2005Sci...309.1833B} {309, 1833}

\bibitem[\protect\citeauthoryear{{Cenko} et~al.,}{{Cenko}
  et~al.}{2006}]{2006ApJ...652..490C}
{Cenko} S.~B.,  et~al., 2006, \mn@doi [\apj] {10.1086/508149}, \href
  {https://ui.adsabs.harvard.edu/abs/2006ApJ...652..490C} {652, 490}

\bibitem[\protect\citeauthoryear{{Chincarini} et~al.,}{{Chincarini}
  et~al.}{2007}]{2007ApJ...671.1903C}
{Chincarini} G.,  et~al., 2007, \mn@doi [\apj] {10.1086/521591}, \href
  {https://ui.adsabs.harvard.edu/abs/2007ApJ...671.1903C} {671, 1903}

\bibitem[\protect\citeauthoryear{{Chincarini} et~al.,}{{Chincarini}
  et~al.}{2010}]{2010MNRAS.406.2113C}
{Chincarini} G.,  et~al., 2010, \mn@doi [\mnras]
  {10.1111/j.1365-2966.2010.17037.x}, \href
  {https://ui.adsabs.harvard.edu/abs/2010MNRAS.406.2113C} {406, 2113}

\bibitem[\protect\citeauthoryear{{D'Elia} et~al.,}{{D'Elia}
  et~al.}{2018}]{2018A&A...619A..66D}
{D'Elia} V.,  et~al., 2018, \mn@doi [\aap] {10.1051/0004-6361/201833847}, \href
  {https://ui.adsabs.harvard.edu/abs/2018A&A...619A..66D} {619, A66}

\bibitem[\protect\citeauthoryear{{Dai}, {Wang}, {Wu}  \& {Zhang}}{{Dai}
  et~al.}{2006}]{2006Sci...311.1127D}
{Dai} Z.~G.,  {Wang} X.~Y.,  {Wu} X.~F.,   {Zhang} B.,  2006, \mn@doi [Science]
  {10.1126/science.1123606}, \href
  {https://ui.adsabs.harvard.edu/abs/2006Sci...311.1127D} {311, 1127}

\bibitem[\protect\citeauthoryear{{Daigne} \& {Mochkovitch}}{{Daigne} \&
  {Mochkovitch}}{1998}]{1998MNRAS.296..275D}
{Daigne} F.,  {Mochkovitch} R.,  1998, \mn@doi [\mnras]
  {10.1046/j.1365-8711.1998.01305.x}, \href
  {https://ui.adsabs.harvard.edu/abs/1998MNRAS.296..275D} {296, 275}

\bibitem[\protect\citeauthoryear{{Daigne} \& {Mochkovitch}}{{Daigne} \&
  {Mochkovitch}}{2000}]{2000A&A...358.1157D}
{Daigne} F.,  {Mochkovitch} R.,  2000, \aap, \href
  {https://ui.adsabs.harvard.edu/abs/2000A&A...358.1157D} {358, 1157}

\bibitem[\protect\citeauthoryear{{De Pasquale} et~al.,}{{De Pasquale}
  et~al.}{2011}]{2011GCN.12228....1D}
{De Pasquale} M.,  et~al., 2011, GRB Coordinates Network, \href
  {https://ui.adsabs.harvard.edu/abs/2011GCN.12228....1D} {12228, 1}

\bibitem[\protect\citeauthoryear{{Eichler} \& {Granot}}{{Eichler} \&
  {Granot}}{2006}]{EG2006}
{Eichler} D.,  {Granot} J.,  2006, \mn@doi [\apjl] {10.1086/503667}, \href
  {http://cdsads.u-strasbg.fr/abs/2006ApJ...641L...5E} {641, L5}

\bibitem[\protect\citeauthoryear{{Evans} et~al.,}{{Evans}
  et~al.}{2007}]{2007A&A...469..379E}
{Evans} P.~A.,  et~al., 2007, \mn@doi [\aap] {10.1051/0004-6361:20077530},
  \href {https://ui.adsabs.harvard.edu/abs/2007A&A...469..379E} {469, 379}

\bibitem[\protect\citeauthoryear{{Evans} et~al.,}{{Evans}
  et~al.}{2009}]{2009MNRAS.397.1177E}
{Evans} P.~A.,  et~al., 2009, \mn@doi [\mnras]
  {10.1111/j.1365-2966.2009.14913.x}, \href
  {https://ui.adsabs.harvard.edu/abs/2009MNRAS.397.1177E} {397, 1177}

\bibitem[\protect\citeauthoryear{{Falcone} et~al.,}{{Falcone}
  et~al.}{2006}]{2006ApJ...641.1010F}
{Falcone} A.~D.,  et~al., 2006, \mn@doi [\apj] {10.1086/500655}, \href
  {https://ui.adsabs.harvard.edu/abs/2006ApJ...641.1010F} {641, 1010}

\bibitem[\protect\citeauthoryear{{Falcone} et~al.,}{{Falcone}
  et~al.}{2007}]{2007ApJ...671.1921F}
{Falcone} A.~D.,  et~al., 2007, \mn@doi [\apj] {10.1086/523296}, \href
  {https://ui.adsabs.harvard.edu/abs/2007ApJ...671.1921F} {671, 1921}

\bibitem[\protect\citeauthoryear{{Fan} \& {Wei}}{{Fan} \&
  {Wei}}{2005}]{FanWei2005}
{Fan} Y.~Z.,  {Wei} D.~M.,  2005, \mn@doi [\mnras]
  {10.1111/j.1745-3933.2005.00102.x}, \href
  {https://ui.adsabs.harvard.edu/abs/2005MNRAS.364L..42F} {364, L42}

\bibitem[\protect\citeauthoryear{{Gao}}{{Gao}}{2009}]{2009ApJ...697.1044G}
{Gao} W.-H.,  2009, \mn@doi [\apj] {10.1088/0004-637X/697/2/1044}, \href
  {https://ui.adsabs.harvard.edu/abs/2009ApJ...697.1044G} {697, 1044}

\bibitem[\protect\citeauthoryear{{Gehrels} et~al.,}{{Gehrels}
  et~al.}{2004}]{2004ApJ...611.1005G}
{Gehrels} N.,  et~al., 2004, \mn@doi [\apj] {10.1086/422091}, \href
  {https://ui.adsabs.harvard.edu/abs/2004ApJ...611.1005G} {611, 1005}

\bibitem[\protect\citeauthoryear{{Ghirlanda}, {Nava}  \&
  {Ghisellini}}{{Ghirlanda} et~al.}{2010}]{2010A&A...511A..43G}
{Ghirlanda} G.,  {Nava} L.,   {Ghisellini} G.,  2010, \mn@doi [\aap]
  {10.1051/0004-6361/200913134}, \href
  {https://ui.adsabs.harvard.edu/abs/2010A&A...511A..43G} {511, A43}

\bibitem[\protect\citeauthoryear{{Ghirlanda} et~al.,}{{Ghirlanda}
  et~al.}{2019}]{GSPGY+2019}
{Ghirlanda} G.,  et~al., 2019, \mn@doi [Science] {10.1126/science.aau8815},
  \href {https://ui.adsabs.harvard.edu/abs/2019Sci...363..968G} {363, 968}

\bibitem[\protect\citeauthoryear{{Giannios}}{{Giannios}}{2006}]{2006A&A...455L...5G}
{Giannios} D.,  2006, \mn@doi [\aap] {10.1051/0004-6361:20065578}, \href
  {https://ui.adsabs.harvard.edu/abs/2006A&A...455L...5G} {455, L5}

\bibitem[\protect\citeauthoryear{{Granot} \& {Ramirez-Ruiz}}{{Granot} \&
  {Ramirez-Ruiz}}{2010}]{GR2010}
{Granot} J.,  {Ramirez-Ruiz} E.,  2010, preprint, \href
  {http://adsabs.harvard.edu/abs/2010arXiv1012.5101G} {} (\mn@eprint {arXiv}
  {1012.5101})

\bibitem[\protect\citeauthoryear{{Granot}, {Ramirez-Ruiz}  \& {Perna}}{{Granot}
  et~al.}{2005}]{2005ApJ...630.1003G}
{Granot} J.,  {Ramirez-Ruiz} E.,   {Perna} R.,  2005, \mn@doi [\apj]
  {10.1086/431477}, \href
  {https://ui.adsabs.harvard.edu/abs/2005ApJ...630.1003G} {630, 1003}

\bibitem[\protect\citeauthoryear{{Greiner} et~al.,}{{Greiner}
  et~al.}{2009}]{2009ApJ...693.1912G}
{Greiner} J.,  et~al., 2009, \mn@doi [\apj] {10.1088/0004-637X/693/2/1912},
  \href {https://ui.adsabs.harvard.edu/abs/2009ApJ...693.1912G} {693, 1912}

\bibitem[\protect\citeauthoryear{{Guidorzi}, {Dichiara}, {Frontera},
  {Margutti}, {Baldeschi}  \& {Amati}}{{Guidorzi}
  et~al.}{2015}]{2015ApJ...801...57G}
{Guidorzi} C.,  {Dichiara} S.,  {Frontera} F.,  {Margutti} R.,  {Baldeschi} A.,
    {Amati} L.,  2015, \mn@doi [\apj] {10.1088/0004-637X/801/1/57}, \href
  {https://ui.adsabs.harvard.edu/abs/2015ApJ...801...57G} {801, 57}

\bibitem[\protect\citeauthoryear{{Hakkila}, {Horv{\'a}th}, {Hofesmann}  \&
  {Lesage}}{{Hakkila} et~al.}{2018}]{2018ApJ...855..101H}
{Hakkila} J.,  {Horv{\'a}th} I.,  {Hofesmann} E.,   {Lesage} S.,  2018, \mn@doi
  [\apj] {10.3847/1538-4357/aaac2b}, \href
  {https://ui.adsabs.harvard.edu/abs/2018ApJ...855..101H} {855, 101}

\bibitem[\protect\citeauthoryear{{Hasco{\"e}t}, {Beloborodov}, {Daigne}  \&
  {Mochkovitch}}{{Hasco{\"e}t} et~al.}{2017}]{2017MNRAS.472L..94H}
{Hasco{\"e}t} R.,  {Beloborodov} A.~M.,  {Daigne} F.,   {Mochkovitch} R.,
  2017, \mn@doi [\mnras] {10.1093/mnrasl/slx143}, \href
  {https://ui.adsabs.harvard.edu/abs/2017MNRAS.472L..94H} {472, L94}

\bibitem[\protect\citeauthoryear{{King}, {O'Brien}, {Goad}, {Osborne}, {Olsson}
   \& {Page}}{{King} et~al.}{2005}]{2005ApJ...630L.113K}
{King} A.,  {O'Brien} P.~T.,  {Goad} M.~R.,  {Osborne} J.,  {Olsson} E.,
  {Page} K.,  2005, \mn@doi [\apjl] {10.1086/496881}, \href
  {https://ui.adsabs.harvard.edu/abs/2005ApJ...630L.113K} {630, L113}

\bibitem[\protect\citeauthoryear{{Kobayashi}, {Zhang}, {M{\'e}sz{\'a}ros}  \&
  {Burrows}}{{Kobayashi} et~al.}{2007}]{2007ApJ...655..391K}
{Kobayashi} S.,  {Zhang} B.,  {M{\'e}sz{\'a}ros} P.,   {Burrows} D.,  2007,
  \mn@doi [\apj] {10.1086/510198}, \href
  {https://ui.adsabs.harvard.edu/abs/2007ApJ...655..391K} {655, 391}

\bibitem[\protect\citeauthoryear{{Krimm} et~al.,}{{Krimm}
  et~al.}{2007}]{2007ApJ...665..554K}
{Krimm} H.~A.,  et~al., 2007, \mn@doi [\apj] {10.1086/519019}, \href
  {https://ui.adsabs.harvard.edu/abs/2007ApJ...665..554K} {665, 554}

\bibitem[\protect\citeauthoryear{{Kr{\"u}hler} et~al.,}{{Kr{\"u}hler}
  et~al.}{2009}]{2009ApJ...697..758K}
{Kr{\"u}hler} T.,  et~al., 2009, \mn@doi [\apj] {10.1088/0004-637X/697/1/758},
  \href {https://ui.adsabs.harvard.edu/abs/2009ApJ...697..758K} {697, 758}

\bibitem[\protect\citeauthoryear{{Kumar} \& {Crumley}}{{Kumar} \&
  {Crumley}}{2015}]{KC2015}
{Kumar} P.,  {Crumley} P.,  2015, \mn@doi [\mnras] {10.1093/mnras/stv1696},
  \href {https://ui.adsabs.harvard.edu/abs/2015MNRAS.453.1820K} {453, 1820}

\bibitem[\protect\citeauthoryear{{Kumar} \& {Narayan}}{{Kumar} \&
  {Narayan}}{2009}]{KN2009}
{Kumar} P.,  {Narayan} R.,  2009, \mn@doi [\mnras]
  {10.1111/j.1365-2966.2009.14539.x}, \href
  {https://ui.adsabs.harvard.edu/abs/2009MNRAS.395..472K} {395, 472}

\bibitem[\protect\citeauthoryear{{Lamb}, {Kann}, {Fern{\'a}ndez}, {Mandel},
  {Levan}  \& {Tanvir}}{{Lamb} et~al.}{2021}]{2021MNRAS.506.4163L}
{Lamb} G.~P.,  {Kann} D.~A.,  {Fern{\'a}ndez} J.~J.,  {Mandel} I.,  {Levan}
  A.~J.,   {Tanvir} N.~R.,  2021, \mn@doi [\mnras] {10.1093/mnras/stab2071},
  \href {https://ui.adsabs.harvard.edu/abs/2021MNRAS.506.4163L} {506, 4163}

\bibitem[\protect\citeauthoryear{{Lamberts} \& {Daigne}}{{Lamberts} \&
  {Daigne}}{2018}]{2018MNRAS.474.2813L}
{Lamberts} A.,  {Daigne} F.,  2018, \mn@doi [\mnras] {10.1093/mnras/stx2951},
  \href {https://ui.adsabs.harvard.edu/abs/2018MNRAS.474.2813L} {474, 2813}

\bibitem[\protect\citeauthoryear{{Lazzati} \& {Perna}}{{Lazzati} \&
  {Perna}}{2007}]{LazzatiPerna2007}
{Lazzati} D.,  {Perna} R.,  2007, \mn@doi [\mnras]
  {10.1111/j.1745-3933.2006.00273.x}, \href
  {https://ui.adsabs.harvard.edu/abs/2007MNRAS.375L..46L} {375, L46}

\bibitem[\protect\citeauthoryear{{Li} et~al.,}{{Li}
  et~al.}{2012}]{2012ApJ...758...27L}
{Li} L.,  et~al., 2012, \mn@doi [\apj] {10.1088/0004-637X/758/1/27}, \href
  {https://ui.adsabs.harvard.edu/abs/2012ApJ...758...27L} {758, 27}

\bibitem[\protect\citeauthoryear{{Liang} et~al.,}{{Liang}
  et~al.}{2006}]{2006ApJ...646..351L}
{Liang} E.~W.,  et~al., 2006, \mn@doi [\apj] {10.1086/504684}, \href
  {https://ui.adsabs.harvard.edu/abs/2006ApJ...646..351L} {646, 351}

\bibitem[\protect\citeauthoryear{{Lyutikov} \& {Blandford}}{{Lyutikov} \&
  {Blandford}}{2003}]{LB2003}
{Lyutikov} M.,  {Blandford} R.,  2003, arXiv e-prints, \href
  {https://ui.adsabs.harvard.edu/abs/2003astro.ph.12347L} {pp
  astro--ph/0312347}

\bibitem[\protect\citeauthoryear{{Malesani} et~al.,}{{Malesani}
  et~al.}{2007}]{2007A&A...473...77M}
{Malesani} D.,  et~al., 2007, \mn@doi [\aap] {10.1051/0004-6361:20077868},
  \href {https://ui.adsabs.harvard.edu/abs/2007A&A...473...77M} {473, 77}

\bibitem[\protect\citeauthoryear{{Margutti}, {Guidorzi}, {Chincarini},
  {Bernardini}, {Genet}, {Mao}  \& {Pasotti}}{{Margutti}
  et~al.}{2010}]{2010MNRAS.406.2149M}
{Margutti} R.,  {Guidorzi} C.,  {Chincarini} G.,  {Bernardini} M.~G.,  {Genet}
  F.,  {Mao} J.,   {Pasotti} F.,  2010, \mn@doi [\mnras]
  {10.1111/j.1365-2966.2010.16824.x}, \href
  {https://ui.adsabs.harvard.edu/abs/2010MNRAS.406.2149M} {406, 2149}

\bibitem[\protect\citeauthoryear{{Margutti}, {Bernardini}, {Barniol Duran},
  {Guidorzi}, {Shen}  \& {Chincarini}}{{Margutti}
  et~al.}{2011}]{2011MNRAS.410.1064M}
{Margutti} R.,  {Bernardini} G.,  {Barniol Duran} R.,  {Guidorzi} C.,  {Shen}
  R.~F.,   {Chincarini} G.,  2011, \mn@doi [\mnras]
  {10.1111/j.1365-2966.2010.17504.x}, \href
  {https://ui.adsabs.harvard.edu/abs/2011MNRAS.410.1064M} {410, 1064}

\bibitem[\protect\citeauthoryear{{Matsumoto}, {Nakar}  \& {Piran}}{{Matsumoto}
  et~al.}{2019a}]{MNP2019}
{Matsumoto} T.,  {Nakar} E.,   {Piran} T.,  2019a, \mn@doi [\mnras]
  {10.1093/mnras/sty3200}, \href
  {https://ui.adsabs.harvard.edu/abs/2019MNRAS.483.1247M} {483, 1247}

\bibitem[\protect\citeauthoryear{{Matsumoto}, {Nakar}  \& {Piran}}{{Matsumoto}
  et~al.}{2019b}]{MNP2019b}
{Matsumoto} T.,  {Nakar} E.,   {Piran} T.,  2019b, \mn@doi [\mnras]
  {10.1093/mnras/stz923}, \href
  {https://ui.adsabs.harvard.edu/abs/2019MNRAS.486.1563M} {486, 1563}

\bibitem[\protect\citeauthoryear{{Mooley} et~al.,}{{Mooley}
  et~al.}{2018}]{MDGNH+2018}
{Mooley} K.~P.,  et~al., 2018, \mn@doi [\nat] {10.1038/s41586-018-0486-3},
  \href {https://ui.adsabs.harvard.edu/abs/2018Natur.561..355M} {561, 355}

\bibitem[\protect\citeauthoryear{{Norris}, {Bonnell}, {Kazanas}, {Scargle},
  {Hakkila}  \& {Giblin}}{{Norris} et~al.}{2005}]{2005ApJ...627..324N}
{Norris} J.~P.,  {Bonnell} J.~T.,  {Kazanas} D.,  {Scargle} J.~D.,  {Hakkila}
  J.,   {Giblin} T.~W.,  2005, \mn@doi [\apj] {10.1086/430294}, \href
  {https://ui.adsabs.harvard.edu/abs/2005ApJ...627..324N} {627, 324}

\bibitem[\protect\citeauthoryear{{Nousek} et~al.,}{{Nousek}
  et~al.}{2006}]{NKGPG+2006}
{Nousek} J.~A.,  et~al., 2006, \mn@doi [\apj] {10.1086/500724}, \href
  {http://cdsads.u-strasbg.fr/abs/2006ApJ...642..389N} {642, 389}

\bibitem[\protect\citeauthoryear{{Oganesyan}, {Ascenzi}, {Branchesi},
  {Salafia}, {Dall'Osso}  \& {Ghirlanda}}{{Oganesyan}
  et~al.}{2020}]{2020ApJ...893...88O}
{Oganesyan} G.,  {Ascenzi} S.,  {Branchesi} M.,  {Salafia} O.~S.,  {Dall'Osso}
  S.,   {Ghirlanda} G.,  2020, \mn@doi [\apj] {10.3847/1538-4357/ab8221}, \href
  {https://ui.adsabs.harvard.edu/abs/2020ApJ...893...88O} {893, 88}

\bibitem[\protect\citeauthoryear{{Panaitescu}}{{Panaitescu}}{2008}]{2008MNRAS.383.1143P}
{Panaitescu} A.,  2008, \mn@doi [\mnras] {10.1111/j.1365-2966.2007.12607.x},
  \href {https://ui.adsabs.harvard.edu/abs/2008MNRAS.383.1143P} {383, 1143}

\bibitem[\protect\citeauthoryear{{Peng}, {Hu}, {Xi}, {Wang}, {Lu}, {Liang}  \&
  {Zhang}}{{Peng} et~al.}{2013}]{2013arXiv1302.4876P}
{Peng} F.-k.,  {Hu} Y.-D.,  {Xi} S.-Q.,  {Wang} X.-G.,  {Lu} R.-J.,  {Liang}
  E.-W.,   {Zhang} B.,  2013, arXiv e-prints, \href
  {https://ui.adsabs.harvard.edu/abs/2013arXiv1302.4876P} {p. arXiv:1302.4876}

\bibitem[\protect\citeauthoryear{{Peng}, {Liang}, {Wang}, {Hou}, {Xi}, {Lu},
  {Zhang}  \& {Zhang}}{{Peng} et~al.}{2014}]{2014ApJ...795..155P}
{Peng} F.-K.,  {Liang} E.-W.,  {Wang} X.-Y.,  {Hou} S.-J.,  {Xi} S.-Q.,  {Lu}
  R.-J.,  {Zhang} J.,   {Zhang} B.,  2014, \mn@doi [\apj]
  {10.1088/0004-637X/795/2/155}, \href
  {https://ui.adsabs.harvard.edu/abs/2014ApJ...795..155P} {795, 155}

\bibitem[\protect\citeauthoryear{{Perna}, {Armitage}  \& {Zhang}}{{Perna}
  et~al.}{2006}]{2006ApJ...636L..29P}
{Perna} R.,  {Armitage} P.~J.,   {Zhang} B.,  2006, \mn@doi [\apjl]
  {10.1086/499775}, \href
  {https://ui.adsabs.harvard.edu/abs/2006ApJ...636L..29P} {636, L29}

\bibitem[\protect\citeauthoryear{{Pescalli}, {Ronchi}, {Ghirlanda}  \&
  {Ghisellini}}{{Pescalli} et~al.}{2018}]{2018A&A...615A..80P}
{Pescalli} A.,  {Ronchi} M.,  {Ghirlanda} G.,   {Ghisellini} G.,  2018, \mn@doi
  [\aap] {10.1051/0004-6361/201732270}, \href
  {https://ui.adsabs.harvard.edu/abs/2018A&A...615A..80P} {615, A80}

\bibitem[\protect\citeauthoryear{{Poolakkil} et~al.,}{{Poolakkil}
  et~al.}{2021}]{2021arXiv210313528P}
{Poolakkil} S.,  et~al., 2021, arXiv e-prints, \href
  {https://ui.adsabs.harvard.edu/abs/2021arXiv210313528P} {p. arXiv:2103.13528}

\bibitem[\protect\citeauthoryear{{Ramirez-Ruiz}, {Granot}, {Kouveliotou},
  {Woosley}, {Patel}  \& {Mazzali}}{{Ramirez-Ruiz}
  et~al.}{2005}]{2005ApJ...625L..91R}
{Ramirez-Ruiz} E.,  {Granot} J.,  {Kouveliotou} C.,  {Woosley} S.~E.,  {Patel}
  S.~K.,   {Mazzali} P.~A.,  2005, \mn@doi [\apjl] {10.1086/431237}, \href
  {https://ui.adsabs.harvard.edu/abs/2005ApJ...625L..91R} {625, L91}

\bibitem[\protect\citeauthoryear{{Rees} \& {M{\'e}sz{\'a}ros}}{{Rees} \&
  {M{\'e}sz{\'a}ros}}{2005}]{2005ApJ...628..847R}
{Rees} M.~J.,  {M{\'e}sz{\'a}ros} P.,  2005, \mn@doi [\apj] {10.1086/430818},
  \href {https://ui.adsabs.harvard.edu/abs/2005ApJ...628..847R} {628, 847}

\bibitem[\protect\citeauthoryear{{Salafia}, {Ghisellini}, {Pescalli},
  {Ghirlanda}  \& {Nappo}}{{Salafia} et~al.}{2015}]{2015MNRAS.450.3549S}
{Salafia} O.~S.,  {Ghisellini} G.,  {Pescalli} A.,  {Ghirlanda} G.,   {Nappo}
  F.,  2015, \mn@doi [\mnras] {10.1093/mnras/stv766}, \href
  {https://ui.adsabs.harvard.edu/abs/2015MNRAS.450.3549S} {450, 3549}

\bibitem[\protect\citeauthoryear{{Sonbas}, {MacLachlan}, {Shenoy}, {Dhuga}  \&
  {Parke}}{{Sonbas} et~al.}{2013}]{2013ApJ...767L..28S}
{Sonbas} E.,  {MacLachlan} G.~A.,  {Shenoy} A.,  {Dhuga} K.~S.,   {Parke}
  W.~C.,  2013, \mn@doi [\apjl] {10.1088/2041-8205/767/2/L28}, \href
  {https://ui.adsabs.harvard.edu/abs/2013ApJ...767L..28S} {767, L28}

\bibitem[\protect\citeauthoryear{{Swenson} \& {Roming}}{{Swenson} \&
  {Roming}}{2014}]{2014ApJ...788...30S}
{Swenson} C.~A.,  {Roming} P.~W.~A.,  2014, \mn@doi [\apj]
  {10.1088/0004-637X/788/1/30}, \href
  {https://ui.adsabs.harvard.edu/abs/2014ApJ...788...30S} {788, 30}

\bibitem[\protect\citeauthoryear{{Swenson}, {Roming}, {De Pasquale}  \&
  {Oates}}{{Swenson} et~al.}{2013}]{2013ApJ...774....2S}
{Swenson} C.~A.,  {Roming} P.~W.~A.,  {De Pasquale} M.,   {Oates} S.~R.,  2013,
  \mn@doi [\apj] {10.1088/0004-637X/774/1/2}, \href
  {https://ui.adsabs.harvard.edu/abs/2013ApJ...774....2S} {774, 2}

\bibitem[\protect\citeauthoryear{{Takahashi} \& {Ioka}}{{Takahashi} \&
  {Ioka}}{2021}]{2021MNRAS.501.5746T}
{Takahashi} K.,  {Ioka} K.,  2021, \mn@doi [\mnras] {10.1093/mnras/stab032},
  \href {https://ui.adsabs.harvard.edu/abs/2021MNRAS.501.5746T} {501, 5746}

\bibitem[\protect\citeauthoryear{{Troja}, {Piro}, {Vasileiou}, {Omodei},
  {Burgess}, {Cutini}, {Connaughton}  \& {McEnery}}{{Troja}
  et~al.}{2015}]{2015ApJ...803...10T}
{Troja} E.,  {Piro} L.,  {Vasileiou} V.,  {Omodei} N.,  {Burgess} J.~M.,
  {Cutini} S.,  {Connaughton} V.,   {McEnery} J.~E.,  2015, \mn@doi [\apj]
  {10.1088/0004-637X/803/1/10}, \href
  {https://ui.adsabs.harvard.edu/abs/2015ApJ...803...10T} {803, 10}

\bibitem[\protect\citeauthoryear{{Von Kienlin} et~al.,}{{Von Kienlin}
  et~al.}{2020}]{2020ApJ...893...46V}
{Von Kienlin} A.,  et~al., 2020, \mn@doi [\apj] {10.3847/1538-4357/ab7a18},
  \href {https://ui.adsabs.harvard.edu/abs/2020ApJ...893...46V} {893, 46}

\bibitem[\protect\citeauthoryear{{Yi}, {Xi}, {Yu}, {Wang}, {Mu}, {L{\"u}}  \&
  {Liang}}{{Yi} et~al.}{2016}]{2016ApJS..224...20Y}
{Yi} S.-X.,  {Xi} S.-Q.,  {Yu} H.,  {Wang} F.~Y.,  {Mu} H.-J.,  {L{\"u}} L.-Z.,
    {Liang} E.-W.,  2016, \mn@doi [\apjs] {10.3847/0067-0049/224/2/20}, \href
  {https://ui.adsabs.harvard.edu/abs/2016ApJS..224...20Y} {224, 20}

\bibitem[\protect\citeauthoryear{{Yi}, {Yu}, {Wang}  \& {Dai}}{{Yi}
  et~al.}{2017}]{2017ApJ...844...79Y}
{Yi} S.-X.,  {Yu} H.,  {Wang} F.~Y.,   {Dai} Z.-G.,  2017, \mn@doi [\apj]
  {10.3847/1538-4357/aa7b7b}, \href
  {https://ui.adsabs.harvard.edu/abs/2017ApJ...844...79Y} {844, 79}

\bibitem[\protect\citeauthoryear{{Yi}, {Du}  \& {Liu}}{{Yi}
  et~al.}{2022}]{2021arXiv211101041Y}
{Yi} S.-X.,  {Du} M.,   {Liu} T.,  2022, \mn@doi [\apj]
  {10.3847/1538-4357/ac35e7}, \href
  {https://ui.adsabs.harvard.edu/abs/2022ApJ...924...69Y} {924, 69}

\bibitem[\protect\citeauthoryear{{Yu} \& {Dai}}{{Yu} \&
  {Dai}}{2009}]{2009ApJ...692..133Y}
{Yu} Y.~W.,  {Dai} Z.~G.,  2009, \mn@doi [\apj] {10.1088/0004-637X/692/1/133},
  \href {https://ui.adsabs.harvard.edu/abs/2009ApJ...692..133Y} {692, 133}

\makeatother
\end{thebibliography}

\appendix
\section{Transformation of bolometric luminosity from aligned to misaligned lines of sight}
\label{sec:B}

We derive the transformation of the observed luminosity from a flashing shell from a given line of sight to another.

We start from the definition for the spectral luminosity:
\begin{equation}
L_\nu(t) = \int \frac{\d L_\nu(t)}{\d \Omega}\d \Omega
\end{equation}

Where $\frac{\d L_\nu (t)}{\d \Omega} = \frac{\d E}{\d t \d\nu \d \Omega}$ is the emitted energy per unit time, frequency and emitting region solid angle and we identify observer time and emission time, disregarding light travel time delays. Changing frames from the emitter's frame to the observer's frame, $\d E$ transforms as $D$ and $\d t$ and $\d \nu$ transform as $D^{-1}$, where $D = \frac{1}{\Gamma (1 - \beta \cos \theta)}$ is the Doppler factor.

Therefore, we have:
\begin{equation}
L_\nu(t) = \int D^3 L'_{\nu'}(t) \d \Omega
\end{equation}

Different lines of sight to the source correspond to different segments of the jet dominating the received emission, i.e. the solid angles with the largest Doppler factor among those which radiate. We denote by $\theta_0$ the total angular size of the emitting region and $\delta \theta$ the angular distance between the line of sight and the edge of the emitting region. We therefore have the following regimes:
\begin{itemize}
\item An aligned line of sight ($\delta \theta = 0$): The observed emission is dominated by a ring with $\theta < 1 / \Gamma$, thus $\d \Omega \sim 2\pi \times 1/2\Gamma^2$, the Doppler factor is $D \sim 2 \Gamma$ and:
\begin{equation}
L_\nu^{\rm on} \sim 8 \pi \Gamma L'_{\nu'}
\end{equation}
\item A slightly misaligned line of sight ($1/\Gamma \ll \delta \theta \ll \theta_0$): The flux is dominated by regions within the emitting region that are the most boosted, with $\delta \theta < \theta < 2 \delta \theta$. In addition, the emitting region is limited to a transverse angular size of $\Delta \phi \sim \pi$ of the emitting region which occupies nearly a half plane in the observer's field. Thus $\d \Omega \sim 3 \delta \theta^2 \Delta \phi/2$, the Doppler factor is $D = \frac{1}{\Gamma (1 - \beta \cos \delta \theta)} \sim 2 \Gamma / (1 + \Gamma^2 \delta \theta^2)$ and:
\begin{equation}
L_\nu^{\rm off} \sim \frac{ 8\Gamma^3}{(1 + \Gamma^2 \delta \theta^2)^3} \frac{3 \Delta \phi \delta \theta^2}{2}L'_{\nu'} \sim \frac{\Delta \phi}{2\pi} S^{-2} L_\nu^{\rm on} 
\end{equation}
where we have used $S = \frac{1 - \beta \cos \delta \theta}{1 - \beta} \sim 1 + \Gamma^2 \delta \theta^2$ and $\Gamma \delta \theta \gg 1$. Therefore $L_\nu^{\rm off} / L_\nu^{\rm on} \sim f_{\rm geo} S^{-2}$, with $f_{\rm geo} = \Delta \phi / 2\pi \sim 1/2$.
\item A significantly misaligned line of sight ($\delta \theta > \theta_0$): The whole emitting region has nearly the same Doppler factor and $\Gamma \delta \theta \gg 1$ still holds. Thus $\d \Omega \sim \theta_0^2$ and one finds $L_\nu^{\rm off} / L_\nu^{\rm on} \sim (\Gamma \theta_0)^2 S^{-3}$. This case does not occur in the setup of the present work.
\end{itemize}

Therefore, defining $S = \frac{1 - \beta \cos \delta \theta}{1 - \beta}$, the spectral luminosity transforms as the following for slightly misaligned lines of sight:
\begin{equation}
\frac{L_\nu^{\rm off}}{L_\nu^{\rm on}} \sim f_{\rm geo} S^{-2}
\label{eq:B:lumnu}
\end{equation}

With a similar reasoning, we have for bolometric luminosities:
\begin{equation}
\frac{L^{\rm off}}{L^{\rm on}} \sim f_{\rm geo} S^{-3}
\label{eq:B:lum}
\end{equation}

\section{The contribution of line-of-sight material to the prompt emission of slightly misaligned observers}
\label{sec:C}

In Sec.~\ref{sec:visi}, we adopt power-law structures in the jet material Lorentz factor and isotropic-equivalent prompt phase emitted energy as a function of latitude (Eqs.~\ref{eq:bo} and \ref{eq:ba}), with parameters $\Gamma_j = 250$, $a = 8$ and $b = 6$. Our derivations suppose that the prompt emission---and therefore the ESD flux---detected by the misaligned observer comes from the material down the observer's line of sight. Here, we will show that this is the case for the adopted parameter values.

Adopting the notations $\epsilon_{\rm em, iso}(\theta)$ and $\Gamma(\theta)$ for the material's \textit{emitted} isotropic-equivalent energy and Lorentz factor,  the \textit{observed} isotropic-equivalent energy at a viewing angle of $\theta_v$ is (e.g., Eq.~5 of \citealt{2015MNRAS.450.3549S}):
\begin{equation}
E_{\rm obs, iso}(\theta_v) = \frac{1}{4\pi}\int\frac{\epsilon_{\rm em, iso}(\theta)}{\Gamma(\theta)}\delta(\theta, \phi, \theta_v)^3\d \Omega
\label{eq:ee}
\end{equation}
where the integration runs over the jet structure, with the solid angle differential element $\d \Omega = \sin \theta \d \theta \d \phi$. The $\delta$ is the relativistic Doppler factor:
\begin{equation}
\delta(\theta, \phi, \theta_v) = \frac{1}{\Gamma(\theta)(1 - \beta(\theta)\cos\chi)}
\end{equation}
and $\chi$ is the angular distance between the observer and the material at coordinates $\theta, \phi$:
\begin{equation}
\cos \chi = \cos \theta_v \cos \theta + \sin \theta_v \sin \theta \cos \phi
\end{equation}

\begin{figure*}
\resizebox{\hsize}{!}{\includegraphics[width=\linewidth]{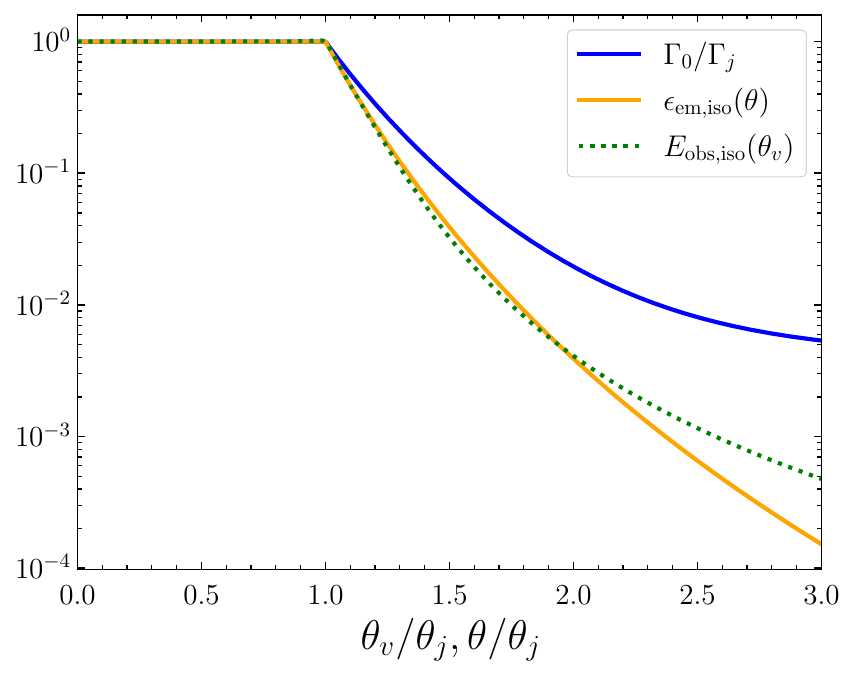}\includegraphics[width=\linewidth]{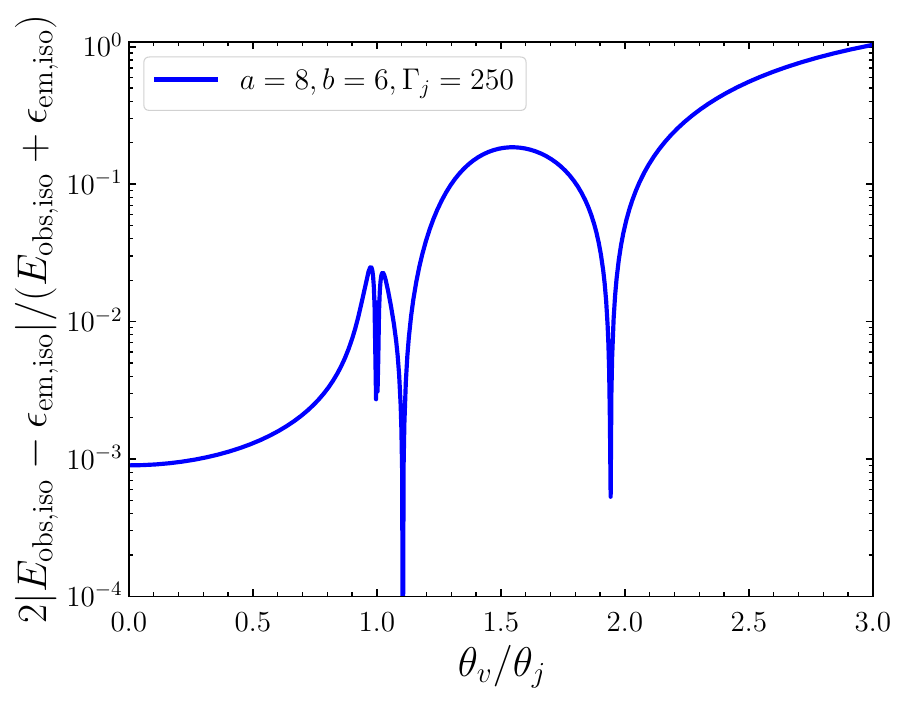}}
\caption{On the same axis for the latitude $\theta$ of the material in the structure and the viewing angle $\theta_v$ of the observer, we plot the prompt emitted energy $\epsilon_{\rm em,iso}(\theta)$ and the observed energy $E_{\rm obs,iso}(\theta_v)$ (left), as well as their relative difference $2|E_{\rm obs} - \epsilon_{\rm em}|/(E_{\rm obs} + \epsilon_{\rm em})$ (right). \new{The relative difference function presents sharp dips whenever the emitted and observed energy functions cross. We can conclude from this figure that}, for the adopted parameters (\new{$\Gamma_j = 250$, $a = 8$ and $b = 6$}, as in the text and the present appendix), the observed energy follows the emitted energy for viewing angles in the slightly misaligned regime, such that the observer's prompt emission can be said to originate from the material on their line of sight.}
\label{fig:structure}
\end{figure*}

In Fig.~\ref{fig:structure} (left; similar to \citealt{BN2019}, Fig.~2 and \citealt{2015MNRAS.450.3549S}, Fig.~4 and following), we plot $\epsilon_{\rm em, iso}(\theta)$ and $E_{\rm obs, iso}(\theta_v)$, normalized to the core emitted energy $\epsilon_{\rm em, iso}(\theta = 0)$, for the above-mentioned parameters and using Eq.~\ref{eq:ee}. In Fig.~\ref{fig:structure} (right), we plot their relative difference, $2 |E_{\rm obs, iso} - \epsilon_{\rm em, iso}|/(E_{\rm obs, iso} + \epsilon_{\rm em, iso})$. \new{This function presents a singularity when the emitted and observed functions cross, as can be seen in the figure.}

We find that for the slightly misaligned regime of viewing angles ($\theta_j \le \theta_v \le 2 \theta_j$), we have $E_{\rm obs, iso}(\theta_v) = \epsilon_{\rm em, iso}(\theta_v)$, to within $\le 18\%$. In other words, the total observed energy at $\theta_v$ equals the emitted energy by the material down the line of sight, at $\theta = \theta_v$, and the prompt emission is dominated by the line-of-sight material. \new{For larger $\theta_v \ge 2\theta_j$, we find that this approximation breaks down.}

\bsp	
\label{lastpage}
\end{document}